\documentclass[aps,twocolumn,groupedaddress,floatfix]{revtex4-2}

\usepackage{graphicx}
\usepackage{hyperref}
\hypersetup{colorlinks=true, linkcolor=blue, anchorcolor=blue, citecolor=blue} %hypertex=true
\graphicspath{{./images/}}
\usepackage{color}
\usepackage{amsmath}
\usepackage{lineno}

\begin{document}

%\linenumbers

%\title{On-chip generation and collective coherent control of multi-partite entangled states}
\title{On-chip generation and collectively coherent control of the superposition of the whole family of Dicke states}

\author{Leizhen Chen$^1$, Liangliang Lu$^{1,2}$, Lijun Xia$^1$, Yanqing Lu$^1$, Shining Zhu$^1$, Xiao-song Ma$^{1,3,4,\ast}$}

\affiliation{
    $^1$National Laboratory of Solid-state Microstructures, School of Physics, College of Engineering and Applied Sciences, Collaborative Innovation Center of Advanced Microstructures, Nanjing University, Nanjing 210093, China\\
    $^2$Key Laboratory of Optoelectronic Technology of Jiangsu Province, School of Physical Science and Technology, Nanjing Normal University, Nanjing 210023, China\\
    $^3$Synergetic Innovation Center of Quantum Information and Quantum Physics, University of Science and Technology of China, Hefei, Anhui 230026, China\\
    $^4$Hefei National Laboratory, Hefei 230088, China\\
	$^{\ast}$e-mails: Xiaosong.Ma@nju.edu.cn}

\date{December 21, 2022}

\begin{abstract}
Integrated quantum photonics has recently emerged as a powerful platform for generating, manipulating, and detecting entangled photons. Multipartite entangled states lie at the heart of the quantum physics and are the key enabling resources for scalable quantum information processing. Dicke state is an important class of genuinely entangled state, which has been systematically studied in the light-matter interactions, quantum state engineering and quantum metrology. Here, by using a silicon photonic chip, we report the generation and collectively coherent control of the entire family of four-photon Dicke states, i.e. with arbitrary excitations. We generate four entangled photons from two microresonators and coherently control them in a linear-optic quantum circuit, in which the nonlinear and linear processing are achieved in a micrometer-scale device. The generated photons are in telecom band, which lays the groundwork for large-scale photonic quantum technologies for multiparty networking and metrology. 
\end{abstract}

\maketitle

\section*{Introduction}
Multipartite quantum states with rich structures are extensively investigated \cite{horodecki2009quantum} and are regarded as core components for implementing quantum information processing tasks. For example, particular multipartite entangled states known as cluster or graph states are universal resources for quantum computation \cite{raussendorf2001one,walther2005experimental,browne2005resource,raussendorf2007topological}. Other states can achieve sub-shot-noise sensitivity in phase estimation, attracting increasing interest in the field of quantum enhanced metrology \cite{leibfried2004toward,pezze2009entanglement,liu2021distributed}. These have triggered high demand for generating and coherent controlling multipartite entanglement. Dicke state \cite{dicke1954coherence} is an important state due to its entanglement being robust against the loss of particles and attractive for practical applications such as multiparty quantum networking \cite{kiesel2007experimental,prevedel2009experimental} and quantum metrology \cite{krischek2011useful}. The Dicke states can also serve as a versatile resource for preparing states of different entanglement classes with lower particle numbers through the projective measurements on individual qubits \cite{kiesel2007experimental}. Note that generalized parity measurements and ancilla qudits can be employed to prepare Dicke states \cite{ionicioiu2008generalized}.

Generally, an N-qubit Dicke state with $m$ excitations is defined as the equal superposition of all basis states and written in the following form:
\begin{equation}
	\left|D_N^m\right>=\frac{1}{\sqrt{C_N^m}}\sum_m P_m\left|0^{N-m}1^m\right>,
	\label{eq:scheme}
\end{equation}
where $\sum_m P_m\left(\cdots\right)$ represents the sum over all possible permutations for $m$ excitations among the $N$ particles. In qubit language, $\left|0^{N-m}1^m\right>$ stands for $m$ qubits in $\left|1\right>$ state, and $N-m$ qubits in $\left|0\right>$ state.

Dicke states have been realized in a variety of physical systems, including photons \cite{kiesel2007experimental,prevedel2009experimental,wieczorek2009experimental,krischek2011useful}, trapped ions \cite{haffner2005scalable}, cold atoms \cite{haas2014entangled,mcconnell2015entanglement,zou2018beating} and superconducting systems \cite{lucero2012computing,mlynek2012demonstrating,wang2020controllable}. 
The past research efforts mainly focused on one or two types of Dicke states, such as those with one or ($N/2$) excitations distributed within $N$ particles.
In particular, photonic systems emerge as a desirable quantum platform owing to the features of weak coupling to the surroundings, individual addressability and high-fidelity qubit operation, opening up ways to systematically study multipartite entanglement. So far, all photonic Dicke states with specific excitations are generated from bulk optical setups with static generation configurations.
The photons generated in these experiments are around 800 nm, which is unsuitable for long-distance transmission because of the signal attenuation in fibers. Photons at telecom wavelength are more applicable to long-distance communications due to the low loss in  fibers and the use of standard high-performance fiber components \cite{cao2019telecom}.
Integrated quantum photonics on silicon, compatible with complementary metal-oxide-semiconductor (CMOS) fabrication, offer intrinsic high optical nonlinearity, dense integration and excellent phase stability and can therefore provide a natural solution for photonic quantum technology. Hundreds of optical components have been integrated onto a single silicon chip, realizing the complex photonic circuits needed to generate and manipulate photon states and achieve large-scale quantum information processing \cite{wang2020integrated,pelucchi2022potential}. Recently, the on-chip generation of two-photon Bell state \cite{silverstone2015qubit}, near-ideal two-photon source \cite{paesani2020near}, four-photon Greenberger-Horne-Zeilinger (GHZ) state \cite{llewellyn2020chip} and cluster state \cite{adcock2019programmable} have been realized with silicon chips. Here, we use integrated photonics to generate the coherent superposition of the entire family of the four-photon Dicke states. Moreover, we harness the advantage of integrated photonics to collectively control all four photons and demonstrate the coherence of various Dicke states. Specifically, we create the path-encoded four-photon Dicke states by using two identical dual Mach-Zehnder interferometer microring (DMZI-ring) photon-pair sources \cite{tison2017path,vernon2017truly,lu2020three,liu2020high}. Field enhancement and independent tuning capabilities of the coupling coefficients of pump, signal and idler photons allow the generation of single photons with high indistinguishability and brightness without using passive filtering. High-fidelity multiphoton operation is realized by linear optics network. All the nonlinear and linear quantum devices are monolithically integrated in silicon. We experimentally observe and coherently control for the first time, to our knowledge, the superposition of the entire family of four-photon Dicke states in the telecom band, which can easily interface with fiber quantum network, implying the flexibility and multiformity of our device.

\section*{Experimental setup}
\begin{figure}[tb]
	\includegraphics[width=0.45\textwidth]{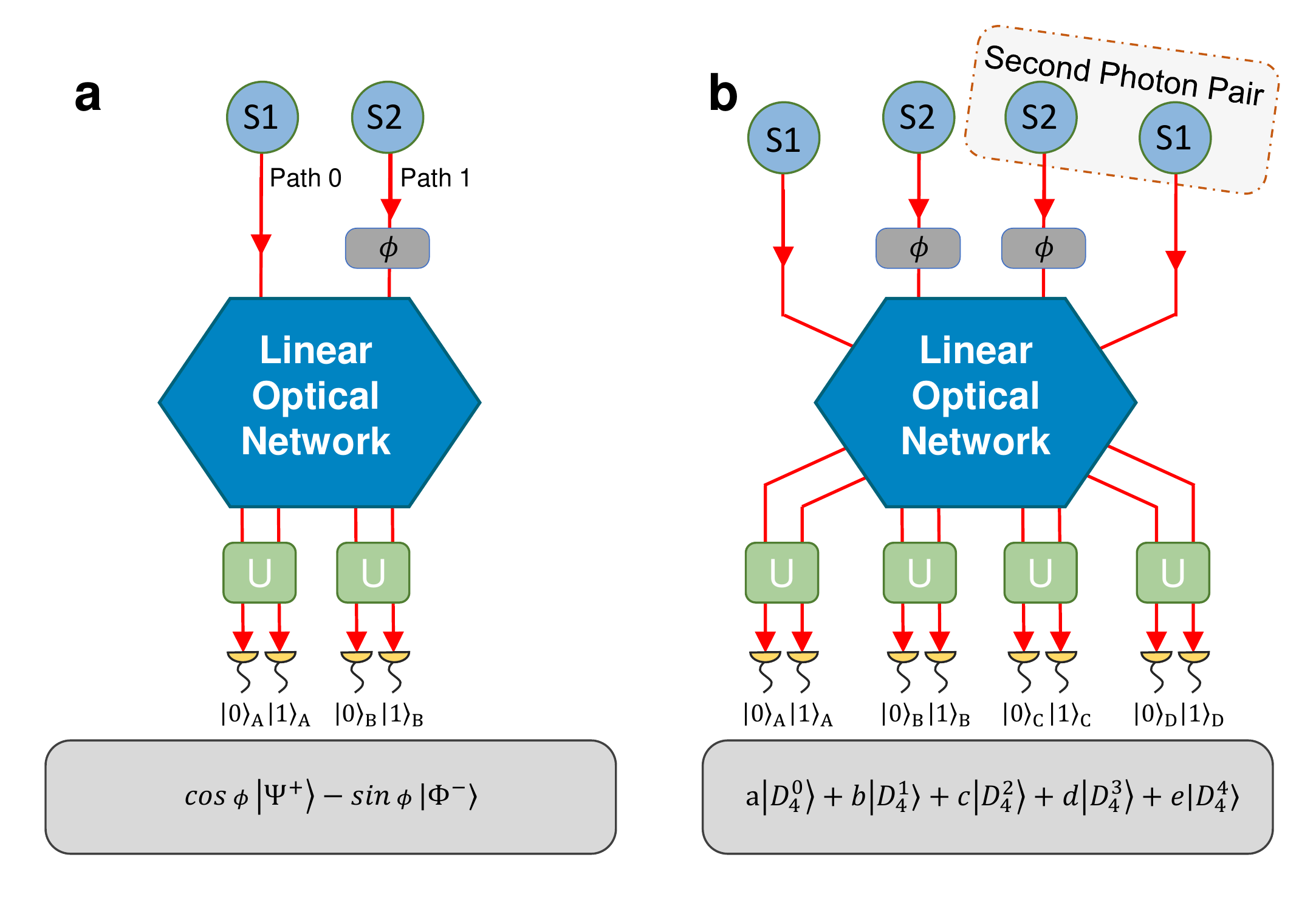} %0.45,0.8
	\caption{\label{fig:scheme} Concept of multiphoton entanglement creation and control. (a) Tunable entangled photon pair generation. A qubit-pair module, consisting of two photon-pair sources (S1 and S2), is coherently pumped. The generated photons are then controlled and combined with beam splitters. The generated quantum state is an entangled two-qubit state after interference on the network, as shown in Eq.(\ref{eq:2photonstate}). This state is then analyzed by universal qubit analyzers (Us) for implementing arbitrary local projective measurements. (b) Tunable entangled four-photon generation. The sources generate two pairs of photons simultaneously. The detections in four detectors result in the generation of the superposition of the whole family of four-photon Dicke states, as shown in Eq.(\ref{eq:4photonstate}). For both the two-photon state and the four-photon state, the phase shift $\phi$ can control all photons coherently and adjust the weights of the respective states.}
\end{figure}
The experimental generation of the multiphoton Dicke state works based on the scheme shown in Fig.\ref{fig:scheme}. Let's start with a two-photon source, made by S1 and S2 [Fig.\ref{fig:scheme}(a)]. We use the following path encoding: S1 could generate two photons in path 0; S2 could generate two photons in path 1. Both sources are coherently pumped by a single laser. After photons in path 1 experience a relative phase $\phi$ to those in path 0, a linear optical network consisting of beam splitters and waveguides combines and splits the path modes 0 and 1. Detecting both photons from the sources, we obtain a two-photon entangled state with the superposition of two Bell states: 
\begin{equation}
	\left|\Psi_2\left(\phi\right)\right>=\cos{\phi}\left|\Psi^+\right>-\sin{\phi}\left|\Phi^-\right>,
	\label{eq:2photonstate}
\end{equation}
where $\left|\Psi^+\right>=1/\sqrt2\left(\left|01\right>+\left|10\right>\right)$ and $\left|\Phi^-\right>=1/\sqrt2\left(\left|00\right>-\left|11\right>\right)$. Note that $\left|01\right>$ stands for $\left|0\right>_A\bigotimes\left|1\right>_B$, in which $\left|0_{A}\right>$ ($\left|1_{A}\right>$) denotes logical value 0 (1) for the dual-rail encoded photonic qubit $A$. In this process, we employ the time-reversed Hong-Ou-Mandel (RHOM) interference between identical photon pairs \cite{chen2007deterministic,silverstone2014chip,silverstone2015qubit}, and Bell-state projection. In this way, the two-photon entangled state and its coherent superposition are created and are controlled via phase $\phi$.

We further consider the probability of two pairs of photons generated at the same time [Fig.\ref{fig:scheme}(b)]. A four-photon entangled state is obtained  when four detectors in each of the four modes respond simultaneously \cite{kiesel2007experimental}. After the linear optical network, we obtain a four-qubit entangled state with the superposition of the whole family of four-photon Dicke states:
\begin{eqnarray}
    \left|\Psi_4\left(\phi\right)\right>&=&\frac{1}{2\sqrt{6}}\left\{	3\sin^2{\phi}\left|D_4^0\right>-6\sin{\phi}\cos{\phi}\left|D_4^1\right>\right.\nonumber\\
    &&+\left.\sqrt6\left(3\cos^2{\phi}-1\right)\left|D_4^2\right>+6\sin{\phi}\cos{\phi}\left|D_4^3\right>\right.\nonumber\\
	&&+\left.3\sin^2{\phi}\left|D_4^4\right>\right\},
	\label{eq:4photonstate}
\end{eqnarray}
By adjusting the relative phase $\phi$, we create various superpositions of all five four-photon Dicke states. Note that the equally weighted superposition of $\left|D_4^0\right>$ and $\left|D_4^4\right>$ corresponds to a GHZ state, and $\left|D_4^1\right>$ and $\left|D_4^3\right>$ are W states \cite{PhysRevLett.92.077901}, respectively. Postselection of one photon per dual-rail qubit plays an important role in this device, and indeed is the source of the measured entanglement. The theoretical postselect efficiency of the four-photon state is 3/32. The details on the evolution and the postselection for the two- and four-photon states, and the scalability of this multipartite entangled state with higher photon numbers can be found in the Supplemental Material. 

\begin{figure*}[htbp]
	\includegraphics[width=0.9\textwidth]{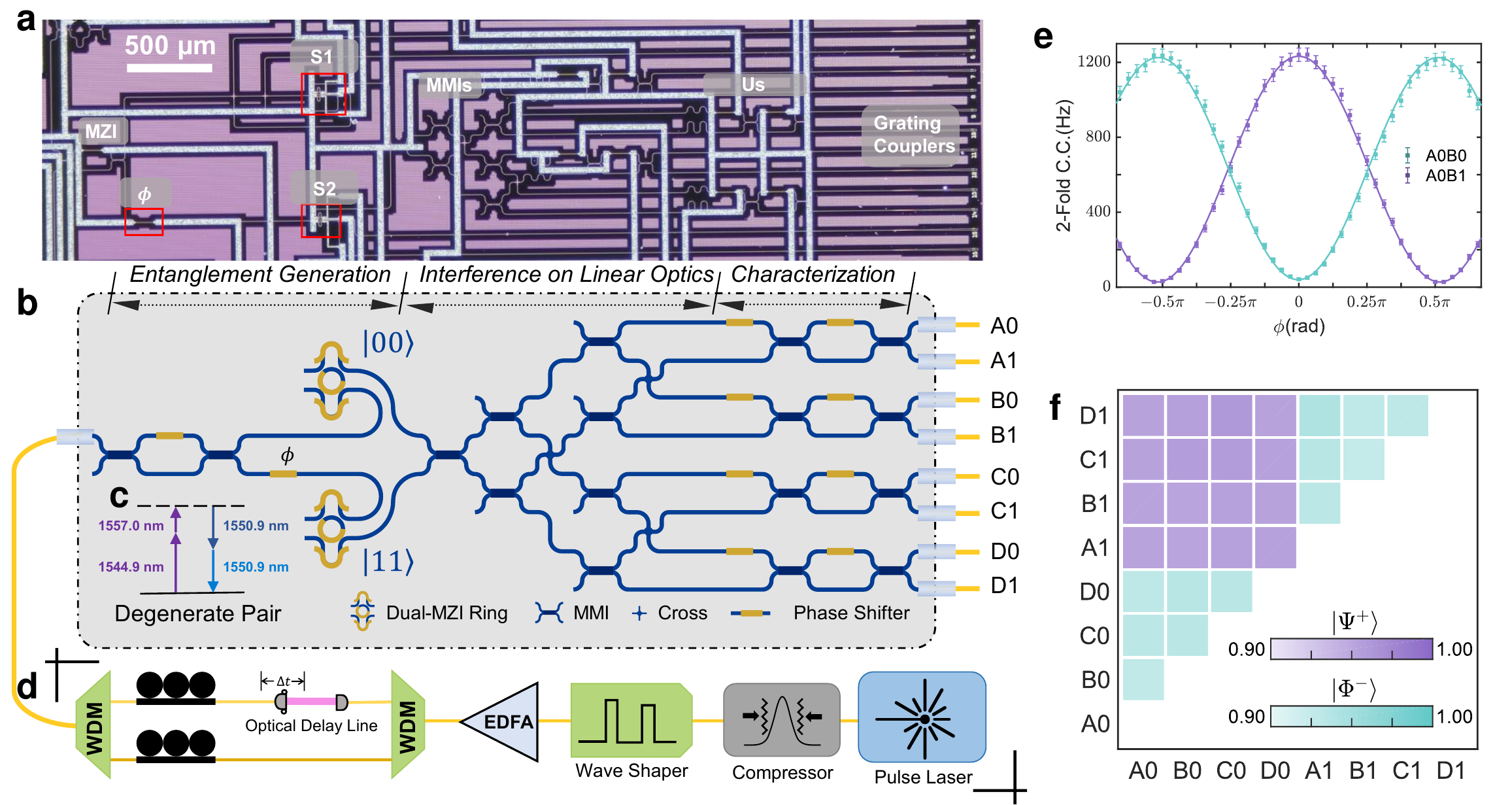}
	\caption{\label{fig:exp} Device description and experimental setup. (a) Photograph of the silicon photonic chip. Important on-chip elements are labeled: two DMZI-rings photon-pair sources (S1 and S2), multimode interference devices (MMIs), universal qubit analyzers (Us) and the phase shifter for coherent control ($\phi$). (b) Experimental setup for generating and characterizing the whole family of Dicke states. Two photon pairs are created. Each pair is in a superposition between two coherently pumped DMZI-ring sources (S1 and S2), and routed by a linear optical network of MMIs to four Us, which are composed of 8 MMIs and 8 phase shifters (PSs). All photons are then coupled out from the chip, filtered and detected by 8 grating couplers, filters and superconducting nanowire single-photon detectors (not shown). All on-chip PSs are controlled with current sources. (c) Two different pump photons (1557.0 nm and 1544.9 nm) generate two identical photons (1550.9 nm) via nondegenerate SFWM process. (d) Dual-color pulse pumping setup to generate degenerate photon pair. A picosecond pump pulse is compressed to increase the bandwidth to over 20 nm, then selected by a wave shaper and amplified by an erbium-doped fiber amplifier (EDFA). After synchronized by WDMs and a tunable optical delay line, two color pulses are coupled into the chip. (e) Twofold coincidence counts between detectors A0 and B1 (corresponding to the projection onto $\left|\Psi^+\right>$, visibility=$95.67\%\pm0.26\%$, purple) and A0 and B0 (corresponding to the projection onto $\left|\Phi^-\right>$, visibility=$93.42\%\pm0.32\%$, green) show the high-quality coherent control of two-photon states, $\left|\Psi_2\left(\phi\right)\right>$. (f) All 28 possible two-photon interference visibilities show the high quality and high homogeneity of our work. The average measured visibilities are $95.87\%\pm0.07\%$ for $\left|\Psi^+\right>$ (purple) and $93.76\%\pm0.09\%$ for $\left|\Phi^-\right>$ (green). Uncertainties are derived from Poissonian statistics and error propagation. }
\end{figure*}

The optical microscope image and schematics of our silicon photonic chip are shown in Figs.\ref{fig:exp}(a) and (b), respectively. To obtain the multiphoton entangled state, identical photons are required. We generate frequency-degenerate photon pairs via the spontaneous four wave mixing (SFWM) [Fig.\ref{fig:exp}(c)] by using a dual-color pulsed pump system [Fig.\ref{fig:exp}(d)] \cite{paesani2019generation,feng2019generation}. A pump laser generates picosecond pulses, whose pulse width is further broadened by a compressor to over 20 nm. Subsequently, a wave shaper filters the incoming broadband pump and select the dual-color pump pulses (1557.0 nm and 1544.9 nm, 40 GHz bandwidth each), which are further amplified by an erbium-doped fiber amplifier (EDFA). A wavelength-division-multiplexing (WDM) network enables us to control the two wavelength pulses independently. In this part, frequency noise suppression, polarization control, and time synchronization are accomplished by WDMs, polarization controllers (PCs), and an optical delay line (ODL), respectively. See Supplemental Material for details on pump control.

On the silicon chip, we use two DMZI-rings \cite{tison2017path,vernon2017truly,lu2020three,liu2020high} as the efficient photon-pair sources. Using these DMZI-rings, we achieve critically coupling for pump light and overcoupling for signal and idler photons at the same time. By doing so, we enhance the use of pump, and achieve a high extraction rate of generated photons (see more experimental details in Supplemental Material). Considering the second-order SFWM process, in which two pairs of photons (four photons) are generated at the same time, we cannot distinguish where these two pairs are generated. They can either be generated from one DMZI-ring or one pair from each DMZI-ring. The two photon pairs are then routed into a linear optical network consisted of seven cascade multimode interference devices (MMIs), acting as balanced beam splitters. At the first MMI, the photon pairs interfere via RHOM \cite{chen2007deterministic,silverstone2014chip,silverstone2015qubit}. Then the remaining MMIs further separate them into four out ports A, B, C and D. Each port has two logic outcomes. For instance, outputs A0 and A1 in Fig.\ref{fig:exp}(b) correspond to the projection of photons into $\left|0\right>_A$ and $\left|1\right>_A$, respectively. At the characterization stage, each qubit is analyzed using a universal qubit analyzer, which is composed of a phase shifter (PS) and a tunable MZI. Then all photons are coupled out from the chip, filtered by off-chip WDMs, and detected with superconducting nanowire single-photon detectors. 

\section*{Results}
First, we verify the indistinguishability of two DMZI-ring photon-pair sources via RHOM interference \cite{chen2007deterministic,silverstone2014chip,silverstone2015qubit}, which is the prerequisite for high-quality multi-photon-state generation. Each two-photon output offers a RHOM interference fringe when scanning the relative phase $\phi$. The coincidence counts with the same path mode (such as A0 and B0) correspond to photon pair in $\left|\Phi^-\right>$ (green curve in Fig.\ref{fig:exp}(e)); the coincidence counts with different path modes (such as A0 and B1) show the complemented results, corresponding to the photon pair in $\left|\Psi^+\right>$ (purple curve in Fig.\ref{fig:exp}(e)). We have thoroughly investigated all 28 possible curves and show their visibility results in Fig.\ref{fig:exp}(f). The average measured visibilities are $95.87\%\pm0.07\%$ for $\left|\Psi^+\right>$ state and $93.76\%\pm0.07\%$ for $\left|\Phi^-\right>$ state. These high visibility fringes indicate the high quality of spectral overlap and qubit entanglement.The deviation of visibilities is mainly due to multiphoton generation events. The higher accidental coincidence from the frequency-degenerate photons may account for the lower visibility of $\left|\Phi^-\right>$ state (see analysis in the Supplemental Material).

\begin{figure}[htbp]
	\includegraphics[width=0.48\textwidth]{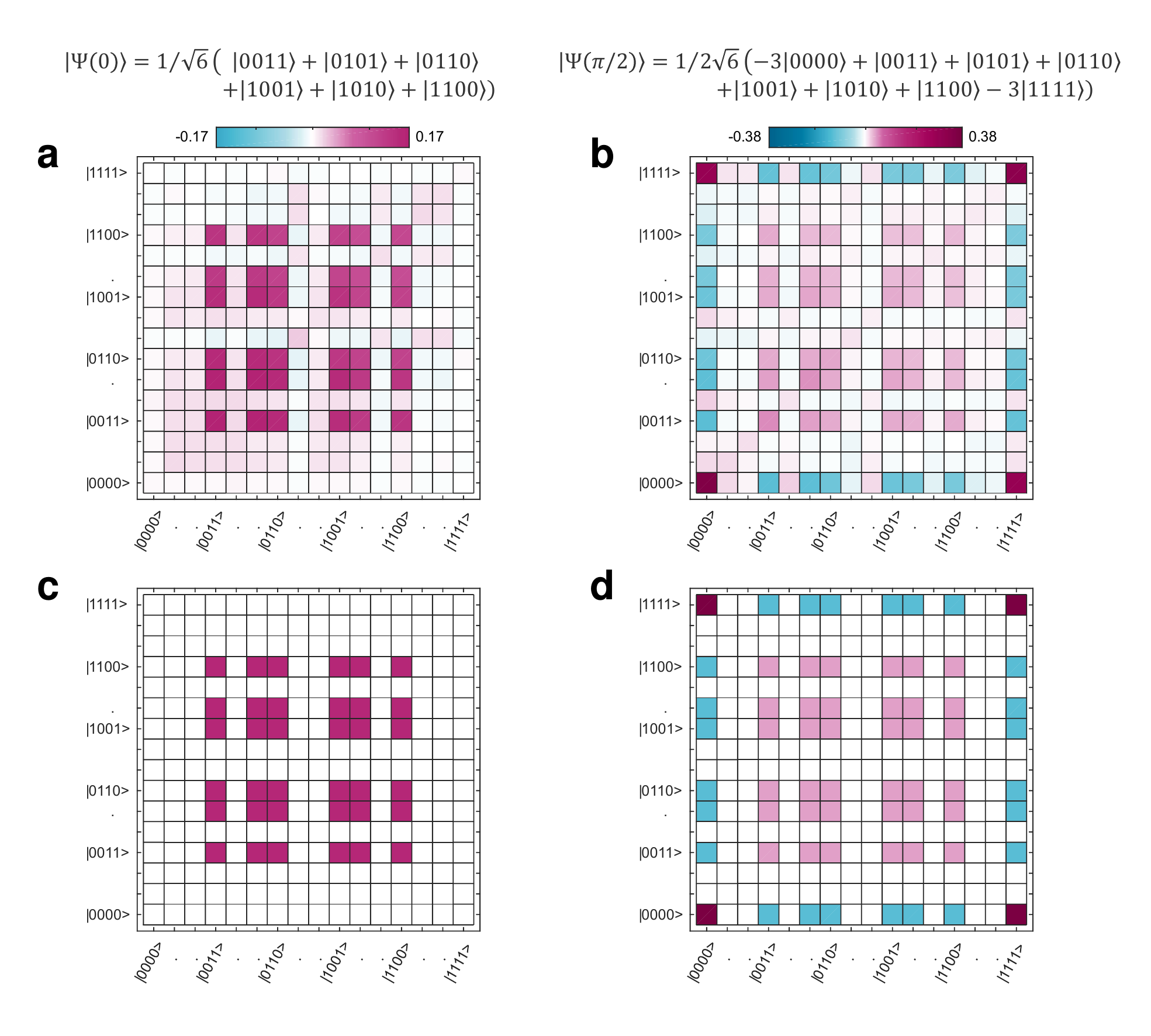}%0.48,0.8
	\caption{\label{fig:tomo} Real part of the density matrix for different states. (a) The symmetric Dicke state $\left|D_4^2\right>$ by setting $\phi=0$. (b) The superposition of four-photon GHZ state $\left|GHZ_4\right>$ and symmetric Dicke state $\left|D_4^2\right>$ by setting $\phi=\pi/2$. The experimentally obtained quantum-state fidelities are $0.817\pm0.003$ for $\left|\Psi\left(0\right)\right>$ and $0.829\pm0.003$ for $\left|\Psi\left(\pi/2\right)\right>$, respectively. The ideal density matrix for $\left|\Psi\left(0\right)\right>$ and $\left|\Psi\left(\pi/2\right)\right>$ are shown in (c) and (d), respectively. Uncertainties are obtained from 100 Monte Carlo simulations with counting Poisson statistics.}
\end{figure}

Having established a high-quality two-photon qubit source, we next investigate the four-photon state when there are two pairs of photons generated in the same pulse. The estimated pair-generation rate is $\sim0.003$ per pulse with $\sim1.3$ mW pulse pump laser coupled onto the chip (a factor of about 500 less than those used in bulk optical experiment \cite{kiesel2007experimental}). To verify the state quality, we choose two special cases of $\left|\Psi\left(\phi\right)\right\rangle$ and characterize them via complete quantum state tomography \cite{james2001measurement}: setting $\phi=0$ results in a symmetric Dicke state $\left|\Psi\left(0\right)\right>=\left|D_4^2\right>$; setting $\phi=\pi/2$ results in a four-photon state with the superposition of GHZ state and symmetric Dicke state: $\left|\Psi\left(\pi/2\right)\right>=1/2\sqrt6 \left(-3\left|0000\right>+\left|0011\right>+\left|0101\right>+\left|0110\right>+\left|1001\right>\right.+\left.\left|1010\right>+\left|1100\right>-3\left|1111\right>\right)$. 
We use 81 settings of all possible combinations of three Pauli bases applied on each photon: $\left\{\left|0\right>,\left|1\right>\right\}$,$\left\{\left|+\right>,\left|-\right>\right\}$, and $\left\{\left|L\right>,\left|R\right>\right\}$, where $\left|+/-\right>=1/\sqrt2\left(\left|0\right>\pm\left|1\right>\right)$ and $\left|L/R\right>=1/\sqrt2\left(\left|0\right>\pm i\left|1\right>\right)$. The tomography measurement takes approximately 3 h per setting and we obtain a total fourfold count of about 1000 for each setting. Figure \ref{fig:tomo} displays the real part of measured and ideal density matrices for $\left|\Psi\left(0\right)\right>$ and $\left|\Psi\left(\pi/2\right)\right>$, respectively. The imaginary part is negligibly small. From experimentally obtained density matrices ($\rho_{\text{exp}}$), we estimate the four-photon state quality with state fidelity to the ideal state ($\rho_{\text{ideal}}$) and obtain the fidelities of $0.817\pm0.003$ for $\left|\Psi\left(0\right)\right>$and $0.829\pm0.003$ for $\left|\Psi\left(\pi/2\right)\right>$, respectively. Here, the fidelity is defined as $F=\text{Tr}\left(\rho_{\text{exp}}\rho_{\text{ideal}}\right)$, and the uncertainty is obtained from Monte Carlo simulation with Poisson statistics. These results show the high quality of our multiphoton source maintains when changing the relative phase $\phi$, indicating that various multiphoton entangled states can be generated on one chip with coherent control. For the symmetric Dicke state, $\left|D_4^2\right>$, we perform the single-qubit projective measurements in different bases and show the conversion between GHZ and W states. We also measure the singlet fraction of the state, showing high robustness against loss of photons and its potential in quantum networking. See Supplemental Material for details.

\begin{figure}[htbp]
	\includegraphics[width=0.48\textwidth]{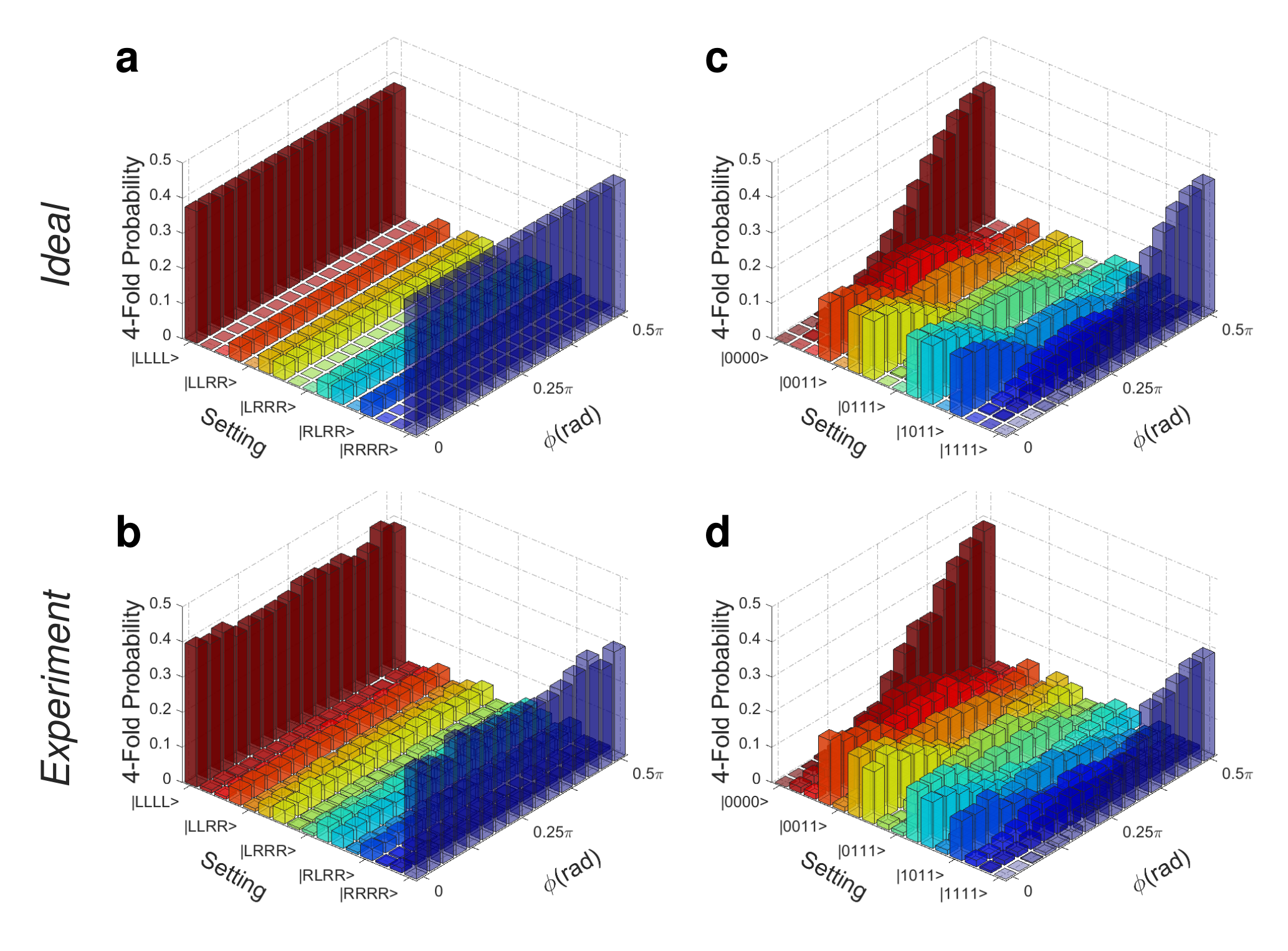} %0.48,0.8
	\caption{\label{fig:statecontrol} Collectively coherent control of the four-photon state. (a), (b) The expected ideal and experimentally measured coincidence distributions for photons measured in the mutually unbiased basis of $\left\{\left|L\right>,\left|R\right>\right\}$. The distributions remain constant independently of the relative phase $\phi$. (c), (d) The ideal and experimental cases for photons measured in the computational basis of $\left\{\left|0\right>,\left|1\right>\right\}$. The relative pump phase $\phi$ rotates the whole family of Dicke state coherently.}
\end{figure}

Unlike the two-photon Bell states where the relative phase $\phi$ can change $\left|\Psi^+\right>$ to $\left|\Phi^-\right>$, it is nontrivial to see what happens to the four photons and how the phase $\phi$ controls all four photons. First, let us rewrite Eq.(\ref{eq:4photonstate}) into the mutually unbiased basis of $\left\{\left|L\right>,\left|R\right>\right\}$:

\begin{eqnarray}
	\left|\Psi_4\left(\phi\right)\right>&=&\frac{1}{2\sqrt6}
	\left(-3e^{-2i\phi}\left|LLLL\right>+\left|LLRR\right>\right.\nonumber\\
     &&+\left.\left|LRLR\right>+\left|LRRL\right>+\left|RLLR\right>\right.\nonumber\\
    &&+\left|RLRL\right>+\left.\left|RRLL\right>-3e^{2i\phi}\left|RRRR\right>\right)
    \label{eq:4photonstateLR}
\end{eqnarray}
which can be viewed as a superposition of a GHZ state and a symmetric Dicke state. One can see that in this basis, the collective phase $\phi$ has no influence on the amplitude of each term. Therefore, when we tune $\phi$ and measure four photons in $\left\{\left|L\right>,\left|R\right>\right\}$ basis, the probabilities of fourfold coincidence stay the same. The theoretical results and the experimental results are shown in Figs.\ref{fig:statecontrol}(a) and (b), respectively, which agree well with each other. In order to prove the coherence of our collective control, we further measure four photons in the $\left\{\left|0\right>,\left|1\right>\right\}$ basis and observe that each of the fourfold coincidences coherently varies with phase $\phi$. We show the theoretical calculations and the experimental results in Figs.\ref{fig:statecontrol}(c) and (d), respectively. To show the agreement between theory and experiment, we calculate the overlap of experimental distribution changes with theoretical cases via the similarity $S=\left(\int\sqrt{\Gamma_{\text{exp}}\Gamma_{\text{ideal}}}\right)^2/\left(\int\Gamma_{\text{exp}}\int\Gamma_{\text{ideal}}\right)$, where $\Gamma_{\text{exp}}$ ($\Gamma_{\text{ideal}}$) is the experimental (ideal) fourfold coincidence distribution changing with phase $\phi$. We obtain the similarity $0.9209\pm0.0015$ for $\left\{\left|L\right>,\left|R\right>\right\}$ basis and $0.9655\pm0.0012$ for $\left\{\left|H\right>,\left|V\right>\right\}$ basis, respectively. 
The experimental results fit well with the expectation, showing the evolution of the whole family of Dicke states. One can further rewrite Eq.(\ref{eq:4photonstateLR}) into the rotated basis $\left\{\left|\theta\right>,\left|\theta_\perp\right>\right\}=\left\{\left(\begin{matrix}\sin{\phi/2}\\\cos{\phi/2}\end{matrix}\right),\left(\begin{matrix}\cos{\phi/2}\\-\sin{\phi/2}\end{matrix}\right) \right\}$
and obtain the symmetric four-photon Dicke state in such basis. 

\section*{Conclusions}
%In conclusion, w
We have presented a single monolithic silicon chip capable of observing the superposition of all five Dicke states with high fidelity by employing resonance-enhanced photon-pair sources. With this device, we have demonstrated the on-chip collectively coherent control of both the two-photon Bell states and the four-photon Dicke states. The experimental setup and methods are generic for observation and coherent control of the entire family of Dicke states with more photon numbers. Such a large-scale integrated quantum circuit can offer the opportunity to generate multiphoton states with larger Hilbert spaces. Although the efficiency to observe states would inevitably decrease with the photon numbers, studying these multiphoton states is important to explore the key techniques of quantum networking, quantum computing, and quantum communication. Our study extends the range of attainable multipartite quantum states in silicon, which has the potential for applications in multiparty quantum networking and quantum enhanced metrology.

\section*{acknowledgments}
	This research was supported by the National Key Research and Development Program of China (Grants No. 2022YFE0137000, No. 2019YFA0308704, and No. 2017YFA0303704), the National Natural Science Foundation of China (Grants No. 11690032 and No. 11321063), the NSFC-BRICS (Grant No. 61961146001), the Leading-Edge Technology Program of Jiangsu Natural Science Foundation (Grant No. BK20192001), the Fundamental Research Funds for the Central Universities, the Innovation Program for Quantum Science and Technology (Grant No. 2021ZD0301500).

 %SI
    \newpage
    \onecolumngrid
    \section*{Supplemental Material}
    \appendix
	\renewcommand*{\appendixname}{}
	\renewcommand*{\thesection}{\arabic{section}}
    \setcounter{figure}{0} 
	\renewcommand*{\thefigure}{S\arabic{figure}}
    \renewcommand*{\thetable}{S\arabic{table}}
    \setcounter{equation}{0}
	\counterwithout{equation}{section}
    \renewcommand*{\theequation}{S\arabic{equation}}

    \section{Details of the silicon device}
    The silicon photonic chip was manufactured using commercial 200 $mm$ wafer platform at the Advanced Micro Foundry (AMF).  The top silicon thickness is 220 $nm$ on the $2 \mu m$ buried oxide (BOX) and covered with $\sim3 \mu m$ silicon dioxide cladding. The waveguides are 500 $nm$ wide using 248 $nm$ lithography for low loss. The minimum feature size of channel waveguides is 0.14 $\mu m$ and the minimum gap between waveguides in our Dicke device is $\sim0.18 \mu m$. To modulate the chip, TiN heaters are placed $\sim2 \mu m$ above the waveguides, using thermo-optic effect to change the phase of waveguides below.

    The insertion loss of on-chip components is important and directly influences the performance of the photonic devices. We measure and estimate the loss of on-chip grating couplers, crosses, and {multimode interferometers (}MMIs{)}.
    
\begin{figure}[bp]
		\includegraphics[width=0.55\textwidth]{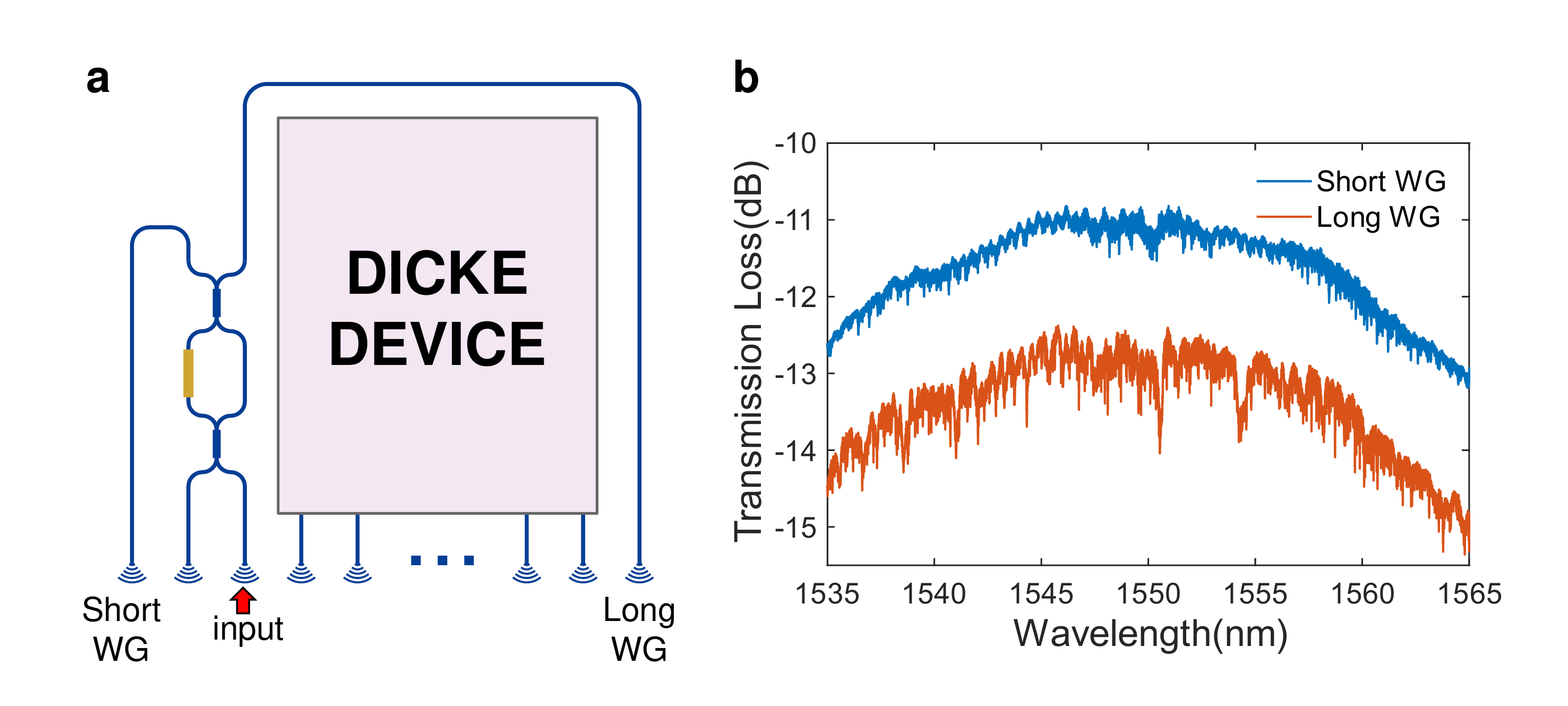}
		\caption{\label{fig:siadd1} (a) Assistant waveguides nearby the Dicke device. The {Mach-Zehnder interferometer (}MZI{)} works as an optical switch. (b) Transmission spectra measured from the short and the long waveguides.}
	\end{figure}

    Grating couplers and waveguide loss: We measure the coupling-assistant waveguides nearby consisting of a testing {Mach-Zehnder interferometer (}MZI{)} structure, a long and a short output waveguides, and grating couplers in and out [Fig.\ref{fig:siadd1}(a)]. The length difference is $\sim$ 10 mm. Assuming that all couplers have the same coupling efficiency, we estimate the propagation loss of waveguide is $\sim1.7$ dB/cm, the typical loss is 2 dB/cm. Considering the insert loss of MZI is $\sim1$ dB, we measure the peak unilateral coupling efficiency is $\sim0.33$ at 1550 $nm$, which means the insert loss of grating coupler is $\sim$4.8 dB/facet.
 
\begin{figure}[tbp]
		\includegraphics[width=0.6\textwidth]{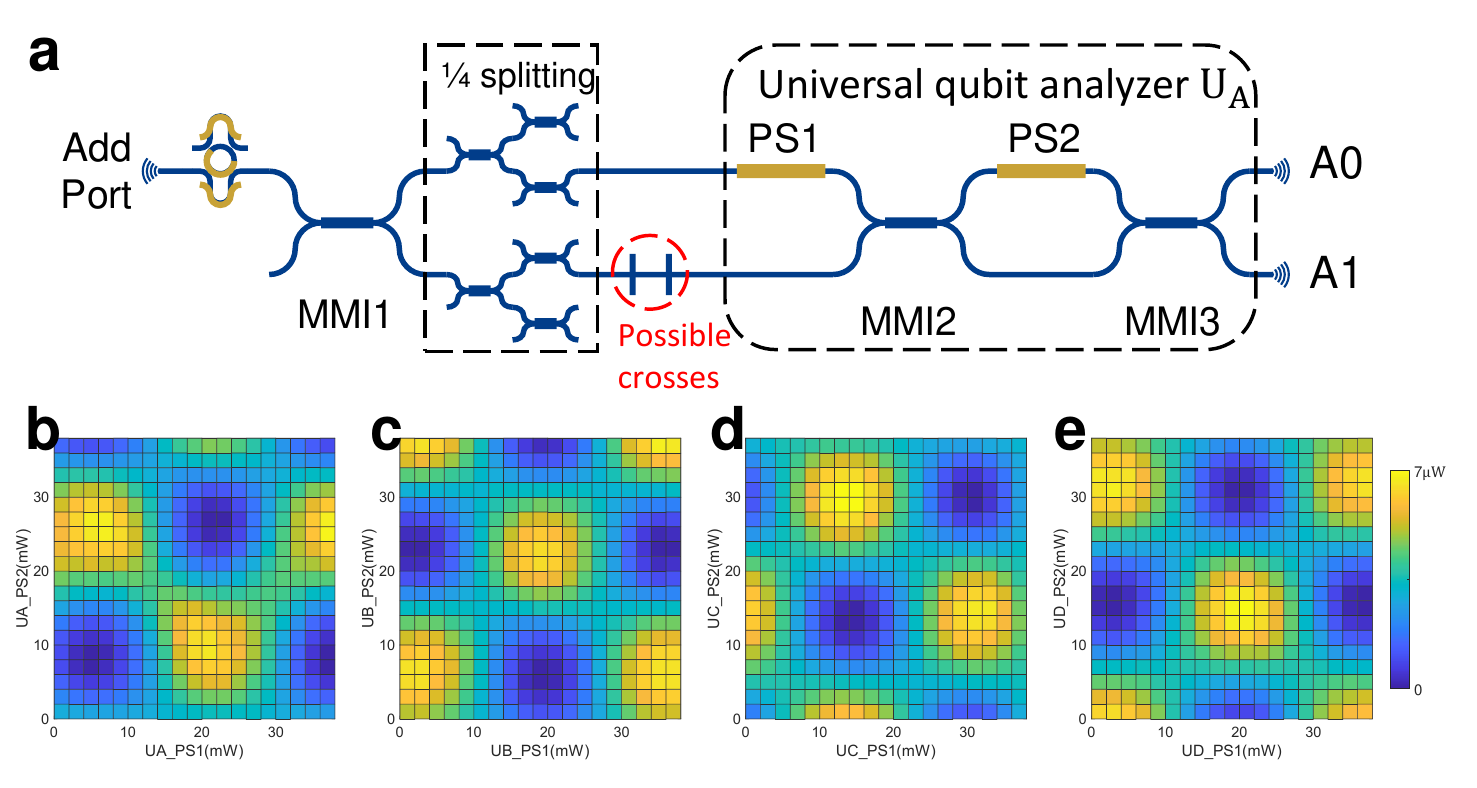}
		\caption{\label{fig:siadd2} (a) The interferometer has 3 {multimode interferometers (}MMIs{)} in the Dicke device. (b)-(e) Transmission curves when scanning PS1 and PS2 in $\rm{U_A - U_D}$.}
	\end{figure}

    Crosses: We estimate the loss of the crosses in the Dicke circuits by interference. We input a laser from the add port of the {dual Mach-Zehnder interferometer microring (}DMZI-ring{)} S1 and detect at the port A0 (B0, C0 or D0) of {universal qubit analyzers (}Us{)} [Fig.\ref{fig:siadd2}(a)]. The circuit with 3 MMIs (MMI1, 2, and 3) constitutes a big interferometer responsed with both the phase shifters PS1 and PS2. Each path between MMI1 and MMI2 has 0-2 crosses introducing loss difference, which can be observed in the interference fringes. Besides, both two paths have 2 additional MMIs introducing the same 6 dB insert loss. Theoretically, the output light can be calculated by the transfer matrixes,
    \begin{equation}
        \text{Out}=(1,0)\cdot \text{MMI}_3 \cdot \text{PS}_2\cdot \text{MMI}_2 \cdot \text{PS}_1 \cdot \text{Loss}\cdot \text{MMI}_1 \cdot (1,0)^{\dagger}
    \end{equation}
    Note that $\text{MMI}_i=\frac{1}{\sqrt{2}}\left(\begin{matrix}1&i\\i&1\\\end{matrix}\right)$, $\text{PS}_j=\left(\begin{matrix}e^{i\varphi_j}&0\\0&1\\\end{matrix}\right)$, $\text{Loss}=\left(\begin{matrix}10^{-0.1\cdot loss}&0\\0&1\\\end{matrix}\right)$. Thus, we obtain the fitting function $f$ describing the big interferometer,  
    \begin{equation}
        f=a\left|10^{-0.1\cdot loss}\left(e^{\varphi_2}e^{\varphi_1}-e^{\varphi_1}\right)-e^{\varphi_2}-1\right|^2,
    \end{equation}  
    where $a$ is the scaling factor to match the measured power. Fig \ref{fig:siadd2} (b) - (e) show the experimental results, from which we also calibrate PS1 and PS2 with the propagating current applied. From the fitting result [Tab.\ref{tab:cross}], we estimate the loss of per cross is 0.2-0.3 dB. The loss is a bit higher because we use our design which was not fully optimized.

    \begin{table}[tbp]
        \centering
        \begin{tabular}{ccccc}
        \hline
             &  A0 &    B0 &    C0 &    D0\\
        \hline
        Fitting loss & -0.59 dB &   0.08 dB &   -0.07dB &   0.42 dB\\
        $n_{cross}$ in path 0 & 0 & 1 & 1 & 2\\
        $n_{cross}$ in path 1 & 2 & 1 & 1 & 0\\
        \hline
        \end{tabular}
        \caption{The fitting loss from 4 big interferometers. The different number of crosses in path 0 and 1 causes a relative loss.}
        \label{tab:cross}
    \end{table}
    
    MMIs: To estimate the loss of MMI, we assume that the loss of other components in the path is known. We input a laser from the add port of a DMZI-ring and measure light at port D0. At the off-resonance frequency, DMZI-ring has no extra loss. The total loss is $\sim$23 dB at 1550 nm including 9 dB from one-by-eight splitting, $\sim$9.6 dB from two couplers, and $\sim1$ dB from waveguide loss. There are 5 MMIs and 3 crosses in the test circuit. Thus, we get the loss of a MMI is 0.5-0.6 dB.
    
\begin{figure}[!htbp]
		\includegraphics[width=0.6\textwidth]{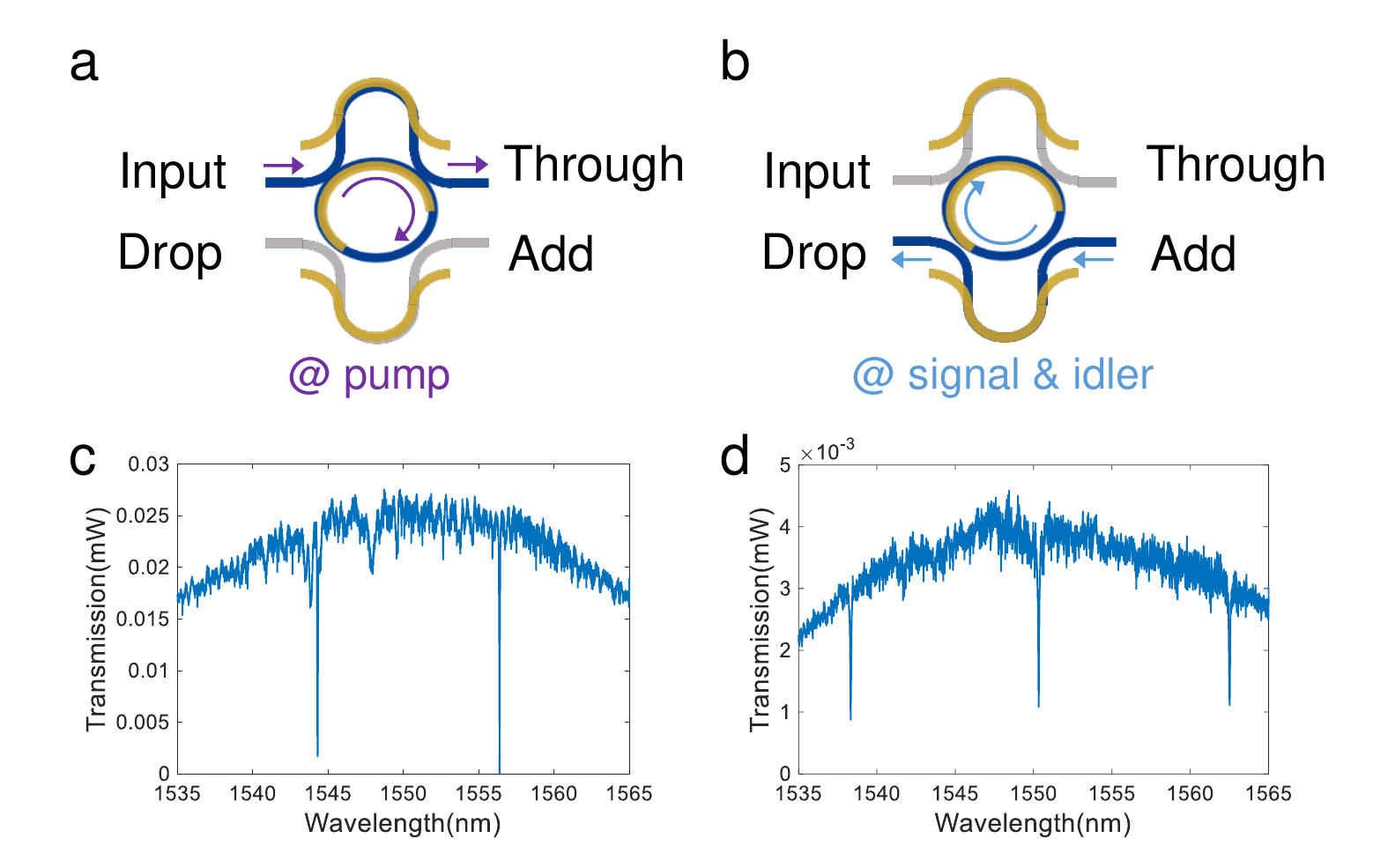}
		\caption{\label{fig:si4} Characterization of DMZI-rings. Working conditions of DMZI-rings show in (a) and (b): (a) For light in pump wavelength, the DMZI-rings are coupled with the input-through bus waveguide and uncoupled with the add-drop bus. (b) For light in signal (idler) wavelength, the DMZI-rings are coupled with the add-drop bus waveguide and uncoupled with the add-drop input-through bus. (c) Transmission spectrum from the input port to the through port. The pump resonance dips show the extinction of $\sim$ 12dB in 1544nm with Q $\sim$ 30800 and $\sim$ 32 dB in 1556nm with Q $\sim$ 36600. (d) Transmission spectrum from the add port to the drop port. The signal (idler) resonance dip shows the extinction of ~5.3dB in 1550nm with Q $\sim$ 17700.}
	\end{figure}

    The DMZI-ring source is designed for photon-pair generation. The DMZI-ring is a four-port device with two bus waveguides. At each bus waveguide, two directional couplers make up an asymmetric Mach-Zehnder interferometer (AMZI). Thus, we can have different coupling conditions for the light at pump and signal (idler) wavelength by designing the length difference of AMZIs and tuning its phase \cite{tison2017path,vernon2017truly,lu2020three,liu2020high}. Under the operating condition, the ring resonator is uncoupled with the add-drop bus waveguide at the pump wavelength and the input-through bus at the signal (idler) wavelength because of the destructive interference. It means the DMZI-ring can be treated as two independent single-MZI ring resonators for pump light [Fig.\ref{fig:si4}(a)] and signal (idler) photons [Fig.\ref{fig:si4}(b)]. Their critical-coupling and overcoupling conditions are decided by the well-designed gap parameters.
	It is worth mentioning that a DMZI-ring offers a high extinction of the input-to-drop light at both pump and signal(idler) wavelength. Thus, we can detect photons mainly from DMZI-rings with minimized influence of photons generated in the waveguides elsewhere on chip, which enhances the signal-to-noise ratio (SNR).
	
	Our DMZI-ring sources is designed with a radius of 15 µm and MZIs’ length difference of $\sim$ 47 µm (half circumference of the ring). The free spectral range (FSR) is $\sim$ 6 nm for ring resonator and $\sim$ 12 nm for both pump and signal (idler) light. The critical-coupling condition gives a high extinction ratio $\sim$ 32 dB for one of pump light [Fig.\ref{fig:si4}(c)], from which we can estimate internal Q factor of the ring resonator. In this case, the loaded Q is equal to the internal Q, and the internal Q is twice the total Q and $\sim$ 73200. Note that the internal Q describes the propagation loss in the ring resonator and the loaded Q describes the coupling to the bus waveguide. We assume that the internal Q is the same for the signal (idler) light, because the wavelength is close. The overcoupling condition gives an extinction ratio $\sim$ 5.3 dB [Fig.\ref{fig:si4}(d)], from which we estimate the loaded Q for signal (idler) light is $\sim$ 23300. Considering the generated photons have the probability to couple out at the add-drop bus or be lost in the ring resonator, the extraction rate of each photon can be defined by the proportion of the total Q to the loaded Q. In our DMZI-rings, the extraction rate of generated photon is $\sim$ 76\% (the extraction rate of ring resonators with critical-coupling condition is 50\%, for comparison).
	Experimentally, we detect a total photon-pair generation rate of $\sim 19000$ Hz from each DMZI-ring source and a coincidence rate of $\sim 300$ Hz with a coincidence-to-accidental ratio of $\sim 45$ because of on-chip one-by-eight splitting of each photon.
 
	\section{Two-qubit entanglement generation via RHOM interference and Bell-state projection}
\begin{figure}[htbp]
		\includegraphics[width=0.7\textwidth]{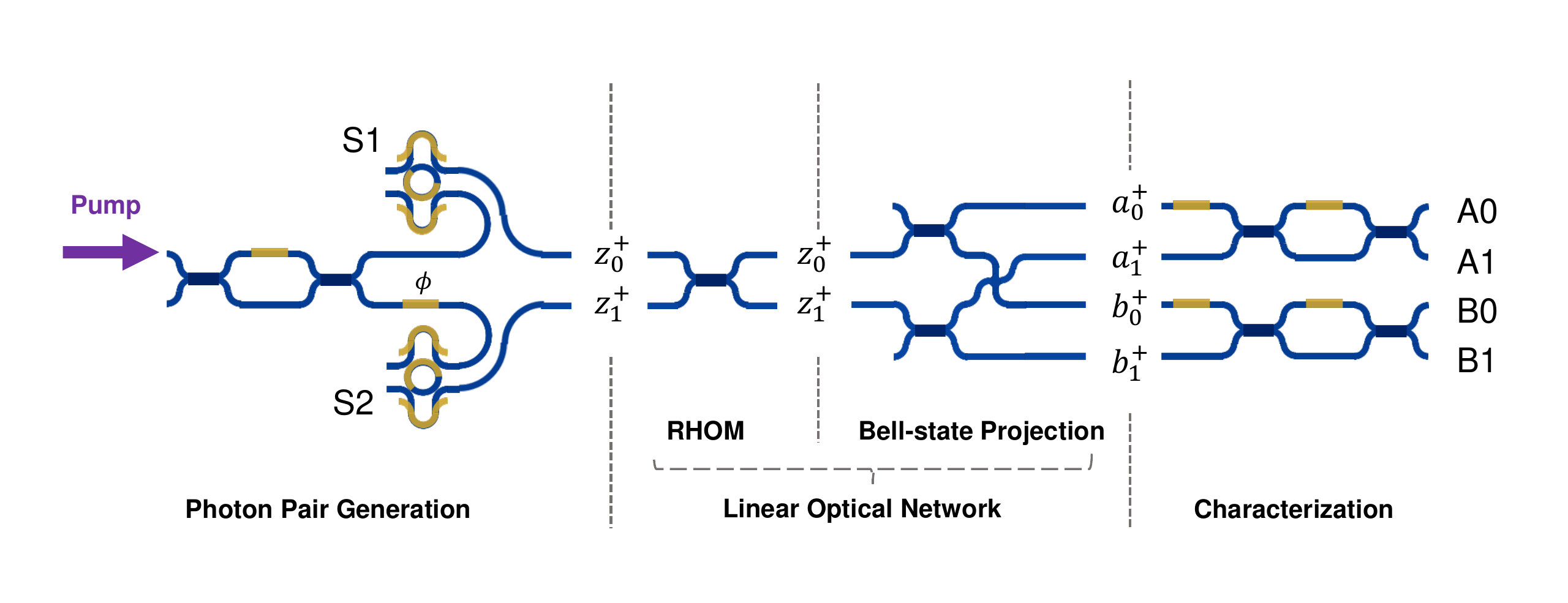}
		\caption{\label{fig:si1} Simplified setup for generating and characterizing the Bell states. A photon pair is generated from Source S1 or S2. Then they are routed into a linear optical network. The first MMI performs the time-reversed Hong-Ou-Mandel (RHOM) interference. The rest of MMIs in the linear optical network separate photons into two outports A and B, which project the photons into the corresponding Bell states. In the characterization stage, universal qubit analyzers measure the photon state.}
	\end{figure}

	Here, we show a detailed derivation of the generation and control of the two-photon state. A simplified linear optical network with fewer MMIs and two outports A and B are shown in {Fig.\ref{fig:si1}}. The derivation can be generalized to larger linear optical networks.  We use creation and annihilation operators to describe the evolution of the photon state.
	
	Let’s start with the photon pair generation part where the phase $\phi$ is included. Two DMZI-ring sources (S1 and S2) are coherently pumped. A pair of photons are generated either from S1 or S2 via the spontaneous four wave mixing (SFWM) process. Balancing the photon-pair generation rate by tuning the phase of the pump’s MZI, the generated two-photon states can be written in the form: 
	\begin{equation}
		1/2\ \left(z_0^+ z_0^++e^{2i\phi}z_1^+ z_1^+\right)\left|\rm{vac}\right>,
		\label{eq:S1-1}
	\end{equation}
	where $z_0^+$ and $z_1^+$ are the creation operators of different path modes from S1 and S2; $\phi$ is the relative phase of the pumps; $\left|\rm{vac}\right>$ is the vacuum state. Then the first MMI at the linear optical network mixes up the photons in two path modes and the time-reversed Hong-Ou-Mandel (RHOM) interference is performed. The generated state is in the form:
	\begin{equation}
		\left[\cos{\phi}\left(z_0^+z_1^+\right)-1/2\ \sin{\phi}\left(z_0^+ z_0^+-z_1^+ z_1^+\right) \right]\left|\rm{vac}\right>.
		\label{eq:S1-2}
	\end{equation}
	Here, we can see that the state is split into three parts: the term $z_0^+z_1^+$ represents the photons have different paths; the terms $z_0^+z_0^+$ and $z_1^+z_1^+ $represent the photons in the same path. The former is further transformed into $\left|\Psi^+\right>$ state in the Bell-state projection part while the superposition of the latter two is further transformed into $\left|\phi\right>$  state. Photons in each mode are separated into two ports A and B with equal probabilities,
	\begin{eqnarray}
		z_0^+\rightarrow1/\sqrt2\ \left(a_0^+ +b_0^+\right),\label{eq:S1-5}\\
		z_1^+\rightarrow1/\sqrt2\ \left(a_1^++b_1^+\right). \label{eq:S1-6}
	\end{eqnarray}
	Note that we move the beam-splitter phase $\pi/2$ into the followed phase shifters in the characterization part, and they can be corrected into zero after initializing .
	Now we have two photons, which may arrive at the same port AA / BB or different ports A and B. If we perform a coincidence detection of photons in the port A and B, the photon state is projected to: 
	\begin{eqnarray}
		z_0^+ z_1^+\rightarrow1/2\ \left(a_0^+ b_1^++a_1^+ b_0^+\right),	\\
		1/2\ \left(z_0^+ z_0^+-z_1^+ z_1^+\right)\rightarrow1/2\ \left(a_0^+ b_0^+ - a_1^+ b_1^+\right),	\\
		1/2\ \left(z_0^+ z_0^++z_1^+ z_1^+\right)\rightarrow1/2\ \left(a_0^+ b_0^+ + a_1^+ b_1^+\right),
	\end{eqnarray}
    where the operator form acts on a vacuum state and get the corresponding Bell state $\left|\Psi^+\right>$, $\left|\Phi^-\right>$, and $\left|\Phi^+\right>$. In this case, the postselect efficiency is 50\%. After Bell-state projection, the final two-photon state is projected into the form as Eq. (2) of the main text, and we can tune the phase $\phi$ to control the weights of $\left|\Psi^+\right>$ and $\left|\Phi^-\right>$.
		
	\section{Expanding the entanglement generation method into four or more photons}
	Based on the principle presented above, we can generalize the method to four or more photon cases by using larger linear optical networks to separate photons into more outports. As SFWM process generates two photons at one time, this scheme only applies to the entangled state with an even number of photons. In such cases, the cascade MMIs after RHOM interference does not serve for a Bell-state projection but a general Dicke-state projection:
	\begin{equation}
		%1/\sqrt{m!(N-m)!}\left(z_0^+\right)^m\left(z_1^+\right)^{N-m}\rightarrow\sqrt{N!/N^N}\left|D_N^m\right>.
		\left(z_0^+\right)^m\left(z_1^+\right)^{N-m}\rightarrow\left|D_N^m\right>.
		\label{eq:S2-1}
	\end{equation}

	Let’s first take the four-photon while-family Dicke state in the main text [Eq.(3)] as an example. 
    %There are two pairs of photons generated in the circuit. Before the Dicke-state projection, these two pairs have not been entangled yet. Thus, after RHOM interference, the four-photon state can be written with the direct product of two two-photon states:
	%\begin{equation}
		%\frac{1}{2} \left[\cos{\phi}\left(z_0^+z_1^+\right)-\frac{1}{2}\sin{\phi}\left(z_0^+z_0^+-z_1^+z_1^+\right)\right]\otimes\left[\cos{\phi}\left(z_0^+z_1^+\right)-\frac{1}{2}\sin{\phi}\left(z_0^+z_0^+-z_1^+z_1^+\right)\right]\left|\rm{vac}\right>.
	%\end{equation}
	%where we ignore the normalization factor.
    In our on-chip circuit, there are two DMZI-ring sources generating photon pairs. At the pair-generation stage, the whole photon state can be written as a squeezed vacuum state,
    \begin{eqnarray}
		\left|\rm{\Psi}\right>&=&\exp{\left[-g\left(z_0^+z_0^++e^{i2\phi}z_1^+z_1^+\right)\right]}\left|\rm{vac}\right>\nonumber\\
        &=&\left\{1-g\left(z_0^+z_0^++e^{i2\phi}z_1^+z_1^+\right)+\frac{1}{2}g^2\otimes^2\left(z_0^+z_0^++e^{i2\phi}z_1^+z_1^+\right)-\frac{1}{3!}g^2\otimes^3\left(z_0^+z_0^++e^{i2\phi}z_1^+z_1^+\right)   \right\}\left|\rm{vac}\right>,
	\end{eqnarray}
    where $g=ie^{i\rm{Arg}(\xi)\tanh{\xi}}$ is referred to as the squeeze parameter; $z_0^+z_0^++e^{i2\phi}z_1^+z_1^+$ is the generation opterator of two coherently pumped DMZI-ring sources; $\phi$ is the relative phase between sources.
    
    Here, we focus on the four-photon (two-pair) term and neglect other terms,
    \begin{equation}
        \frac{1}{2}g^2\otimes^2\left(z_0^+z_0^++e^{i2\phi}z_1^+z_1^+\right)=
        \frac{1}{2}g^2\left(z_0^+z_0^+z_0^+z_0^++e^{i2\phi}z_0^+z_0^+z_1^+z_1^++e^{i4\phi}z_1^+z_1^+z_1^+z_1^+\right).
    \end{equation}
    As can be seen, the two pairs can be generated either in one DMZI-ring or simultaneously one pair in each DMZI-ring. Then the first MMI at the linear optical network mixes up the four photons, and we get:
    
	\begin{eqnarray}
		\left\{\frac{1}{8}\sin^2{\phi}z_0^+z_0^+z_0^+z_0^+
		-\frac{1}{2}\sin{\phi}\cos{\phi}z_0^+z_0^+z_0^+z_1^+ 
		+\frac{1}{4}\left(3\cos^2{\phi}-1\right)z_0^+z_0^+z_1^+z_1^+\right.\ &\nonumber\\
		\left.+\frac{1}{2}\sin{\phi}\cos{\phi}z_0^+z_0^+z_0^+z_1^+
		+\frac{1}{8}\sin^2{\phi}\ z_1^+z_1^+z_1^+z_1^+\right\}&\left|\rm{vac}\right>.
	\end{eqnarray}
	We can see all terms of $\left(z_0^+\right)^m\left(z_1^+\right)^{4-m}$ occur with different weights changing with $\phi$.
	Then the rest of MMIs separated photons equiprobably into four ports, which gives:
	\begin{eqnarray}
		z_0^+\rightarrow1/2\ \left(a_0^++b_0^++c_0^++d_0^+\right),
		\label{eq:S2-4}\\
		z_1^+\rightarrow1/2\ \left(a_1^++b_1^++c_1^++d_1^+\right).
		\label{eq:S2-5}
	\end{eqnarray}
	After selecting the fourfold coincidence, each term of $\left(z_0^+\right)^m\left(z_1^+\right)^{4-m}$ is projected into the corresponding Dicke states, following:
	\begin{eqnarray}
	1/2\sqrt{6}\ z_0^+z_0^+z_0^+z_0^+ &\rightarrow& \sqrt{6}/8 \left|D_4^0\right>, \\
	1/\sqrt{6}\ z_0^+z_0^+z_0^+z_1^+ &\rightarrow& \sqrt{6}/8 \left|D_4^1\right>, \\
	1/2\ z_0^+z_0^+z_1^+z_1^+ &\rightarrow& \sqrt{6}/8 \left|D_4^2\right>, \\
	1/\sqrt{6}\ z_0^+z_1^+z_1^+z_1^+ &\rightarrow& \sqrt{6}/8 \left|D_4^3\right>, \\
	1/2\sqrt{6}\ z_1^+z_1^+z_1^+z_1^+ &\rightarrow& \sqrt{6}/8 \left|D_4^4\right>.
	\end{eqnarray}
	Note that the state on the left side is normalized, and the weight on the right side shows the possibility to observe each Dicke state. Thus, we get the four-photon state with the superposition of the whole family of Dicke states with the corresponding weights:
	\begin{equation}
		\frac{3}{16}\sin^2{\phi}\left|D_4^0\right>-\frac{3}{8}\sin{\phi}\cos{\phi}\left|D_4^1\right>
		+\frac{\sqrt6}{16}\left(3\cos^2{\phi}-1\right)\left|D_4^2\right>+\frac{3}{16}\sin{\phi}\cos{\phi}\left|D_4^3\right>
		+\frac{3}{16}\sin^2{\phi}\left|D_4^4\right>,
		\label{eq:4photonstate_unnormal}
	\end{equation}
	which is the same as Eq.(3) in the main text after normalization. From Eq.\ref{eq:4photonstate_unnormal}, we further calculate the efficiency to observe four-photon state from all four-photon events. The generation efficiency is always 3/32, independent of the phase $\phi$.
	
	If we employ a larger linear optical network with more out ports, we can generate the entangled state with the superposition of the whole family of multiphoton Dicke state. The Dicke-state projection is performed by splitting and sending the photon to each port of the linear optical network. The key point is that photons in mode $z_0^+$ need have the same distribution to each port as photons in mode $z_1^+$. The maximal generation efficiency is $N!/N^N$ and reached when each photon splitting evenly to each port. For the photon number like 6, 10, or 12, photons may arrive at several ports with higher possibility. In this case, the input state can still be projected to corresponding Dicke states but with low efficiency. To get the balanced splitting and maximum utilization of photons, BSs with specific or tunable reflectivity is required.
    
    Recently, an on-chip practical design for symmetric Dicke states $\left|D_N^{N/2}\right>$ is proposed by Ref \cite{zhu2020reconfigurable}, which is based on graph theory and perfect matching \cite{krenn2017quantum,gu2019quantum2,gu2019quantum3}. Their setup with beam-splitter network is similar to our experiment setup. What differs is that by tuning the relative phase of two DMZI-ring sources, we can control the input states before Dicke-state projection and thus obtain the coherent superposition the entire family of the multiphoton Dicke states. Ref \cite{zhu2020reconfigurable} also proposes a different and complex on-chip design for generation of general Dicke states $\left|D_N^{m}\right>$, based on the graph of general Dicke states \cite{gu2019quantum3}. In our case, $\left|D_N^{m}\right>$ can be generated via Dicke-state projection, but it is still challengeable to prepare photons in $\left(z_0^+\right)^m\left(z_1^+\right)^{N-m}$ with on-chip probabilistic photon sources.
      
	\begin{figure}[htbp]
		\includegraphics[width=0.6\textwidth]{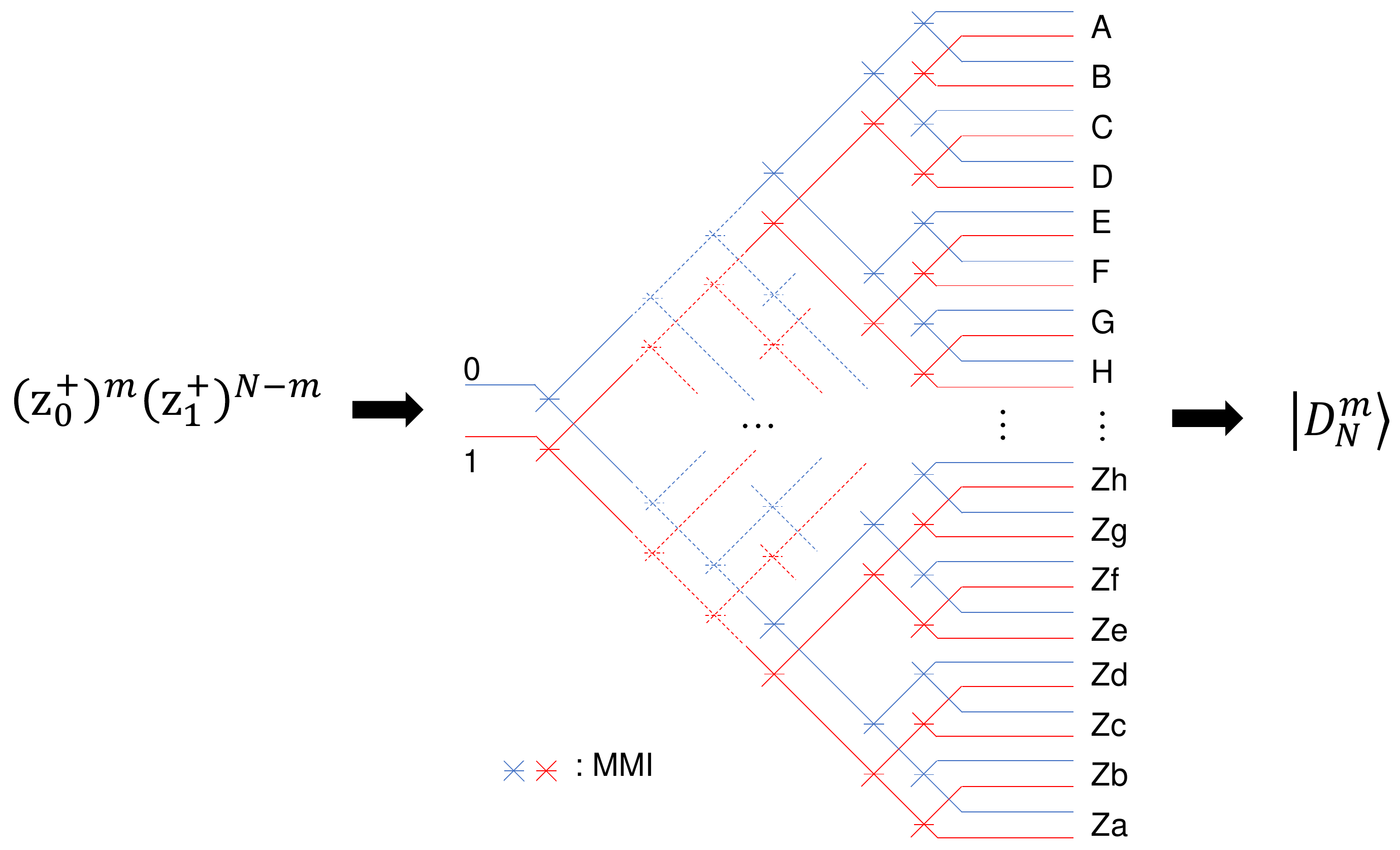}
		\caption{\label{fig:si2} Setup of a scalable linear optical network for multiphoton Dicke-state projection. Blue (red) lines denote photon encoding with path mode 0 (1). The input photon state $\left(z_0^+\right)^m\left(z_1^+\right)^{N-m}$ results in the corresponding state $\left|D_N^m\right>$. A input state with superposition of all $\left(z_0^+\right)^m\left(z_1^+\right)^{N-m}$ — for instance, our source after RHOM as Eq.(\ref{eq:S1-2}) – results in a multiphoton whole-family Dicke-state. The solid lines show a setup for 16-photon Dicke state.}
	\end{figure}	
    
	\section{Dual-color pulsed pump system}
\begin{figure}[htbp]
		\includegraphics[width=0.8\textwidth]{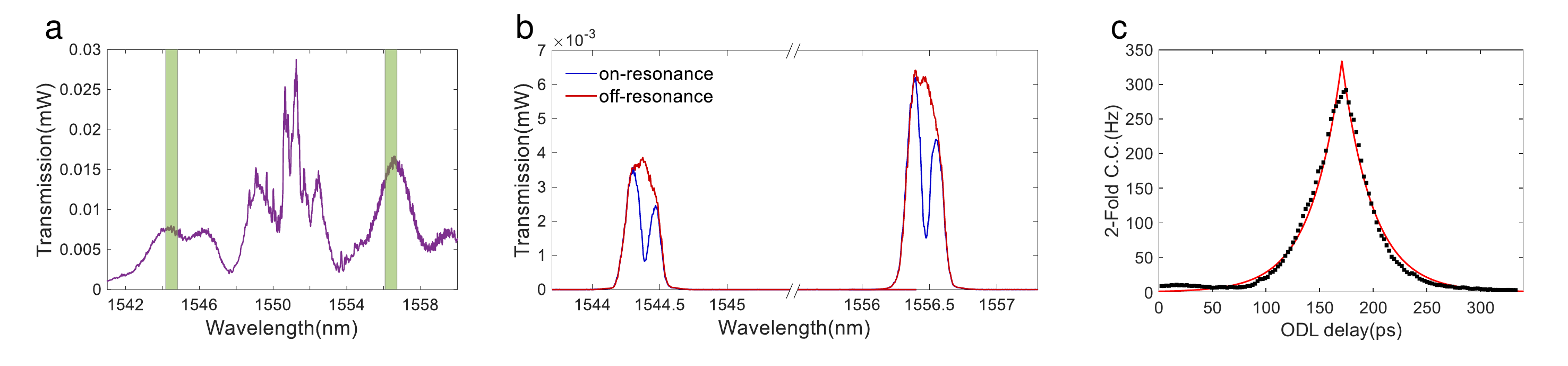}
		\caption{\label{fig:si3} The dual-color pulsed pump system for nondegenerate SFWM. (a) Spectrum of the compressed pulsed pump. With proper pump power, the bandwidth of the output pulse is broadened over 20nm. Green rectangles indicate the selected pump wavelength for generating photon pairs. (b) The transmission spectra of dual-color pulsed pump measured after the chip. The off-resonance (red) and on-resonance (blue) spectra of the transmission are shown respectively. (c) Measurement of degenerate pair generation rate by scanning the delay of the ODL. The overlap of two pulses promotes the nonlinear interaction strength and gives an enhancement in coincidence counts of photon pairs.}
	\end{figure}

	To generate degenerate photon pairs via SFWM process, two pump pulses with different frequency is needed. In our work, we use a dual-color pulsed pump system [Fig.2(d) in main text] to select, filter and synchronize these two pluses. A pulse laser emits pulse light at a repetition frequency of 500 MHz. The pulse’s bandwidth is $\sim$ 1.5 nm, which needs to be broadened in order to cover the two wavelength components we need. A compressor broadens the import light via supercontinuum generation. With the proper power of the input pulse, we obtain the broadened bandwidth over 20 nm with the needed wavelength components just lying on the maximum of two sidebands (shown in two green rectangular areas in Fig.\ref{fig:si3}(a)). After that, we use a wave shaper to select two wavelength components with 40 GHz bandwidth each. These two pulses are then sent into one fiber, and further amplified by an erbium-doped fiber amplifier (EDFA). Considering the different gain factors of two pulses, the wave shaper also reduces one pulse by 5 dB to balance the final two pulses after amplification. Fig.\ref{fig:si3}(b) shows the spectrum of dual-color pulsed pump after the DMZI-rings. When tuning DMZI-ring’s resonance away from the pulsed pump wavelength, the intensity of two pulses is similar (red line). When DMZI-ring matches the pulsed pump, we can see two resonant dips (blue line).

	To get the maximum photon-pair generation rate, two pulses are synchronized via the optical delay line (ODL) in the wavelength-division-multiplexing (WDM) network. WDMs not only filter the amplifier spontaneous-emission noise (ASE) of pulsed pump, but also split and combine two pulses. Thus, the ODL can control the relative arrival time of the two pulses. Scanning the delay of one pulse via the ODL, the photon-pair generation rate reaches the maximum when the two pulses overlaps perfectly [Fig.\ref{fig:si3}(c)]. The overlap can be described with the temporal cross-correlation function of two pump pulses: $G_{p1,p2}^{\left(2\right)}\ \left(\tau\right)\propto\exp{\left(-2\pi\Delta v\left|\tau\right|\right)}$ \cite{ou1999cavity}, where $\Delta v$ is related to the pump resonance linewidth $\sim$ 40 pm (5 GHz), not selected bandwidth of $\sim$ 320 pm (40 GHz), because the SFWM process happens in the DMZI-rings. The {full width at half maximum (}FWHM{)} of the fringe is $\sim$ 40 ps with the fitted $\Delta v \sim$ 5.5 GHz, which is consistent with the linewidth of the DMZI-rings.
	
    \section{The Klyshko efficiency of the dual-pump SFWM process}
    The Klyshko efficiency or the heralding efficiency, $\left(\eta_s\right)_{raw}=CC(s,i)/SC(i)$ \cite{klyshko1980use}, is used to evaluate the  total loss of signal photon from the source to the detector. However,  under dual-color pulsed pump condition, the single-pump SFWM process still happens and also provides one of photons at the same frequency we detected. Thus, subtracting these components, we get the net Klyshko efficiency,
    \begin{equation}
         \left(\eta_s\right)_{net}=\frac{CC(s,i)_{net}}{SC(i)_{net}}=\frac{CC(s,i)-ACC(s,i)}{SC(i)_{pump1\&pump2}-SC(i)_{pump1}-SC(i)_{pump2}}
    \end{equation}      
    where $CC(s,i)$ is the coincidence-count rat, $SC(i)$ is the single-count rate, and $ACC(s,i)$ is the accidental coincidence-count rate. Tab.\ref{tab:SFWM} is the experimental result when we block one of the dual-color pumps or suppress its coupling with polarization rotation. We can see move than half of the single counts are from the single-pump SFWM process. From the single-count rate, we estimate the strength relation of different SFWM processes $g$:$g_1$:$g_2$$\approx$1:1.5:0.86, where $g_1$($g_2$) is the SFWM intensity with single pump at 1545 nm (1557 nm). When tuning the polarization, there are still some light coupled into the chip, which results in a small coincidence rate. From the experimental result, we estimated the net Klyshko efficiency is $\sim$0.024-0.028, which means the total loss $\sim$16 dB from DMZI-rings to detectors.

    \begin{table}[htbp]
        \centering
        \begin{tabular}{ccccc}
        \hline
             &  SC1(Hz) &    SC2(Hz) &    CC(Hz) &    ACC(Hz)\\
        \hline
        Dual-color pump & 112119 & 92293 & 1135 & 18\\
        &&&&\\
        Single pump @1545 nm  & 50908 & 43910 & 35 & 4.8\\
        (suppress coupling of the other pump) &&&&\\
        Single pump @1557 nm & 16864  & 14506  & 14  & 0.5\\
        (suppress coupling of the other pump) &&&&\\
        Single pump @1557 nm  & 1660  & 14231  & 0.5  & 0.5\\
        (block light from ODL)&&&&\\
        \hline
        \end{tabular}
        \caption{The single-count rate and twofold coincidence-count rate measured under different pump conditions. }
        \label{tab:SFWM}
    \end{table}

    \section{The four-photon coincidence rate of the four-photon state}
    Fourfold coincidence rate quantifies the efficiency we produce and detect the quantum state, and it directly determines the integral time of the data collection. We can estimate the four-photon coincidence rate from the measured two-photon coincidence rate. Here we take $\left|D_4^2\right>(\phi=0)$ as an example. Two-photon component is $gz_0^+z_1^+\left|\rm{vac}\right>=g\left|1_0,1_1\right>$, and the pair-generation rate per pulse is $p\propto g^2$. In this case, the generated photon pair is split into one photon in each path 0 and 1. So the two-photon coincidence from one-pair generation is $pf\eta_s\eta_i$, where $f=500$ MHz is the repetition frequency of pulsed laser and $\eta_{s/i}(s/i=A,B,C,D)$ is the loss at each port. Considering the multi-pair generation at a high power of pulsed pump, the single-count rate and two-photon coincidence rate can be described as,
    \begin{eqnarray}
    SC(s)_{net}&=&DC_s+pf\eta_s+p^2 f \cdot2\eta_s+O(p^3), \\
    SC(i)_{net}&=&DC_i+pf\eta_s+p^2 f \cdot2\eta_i+O(p^3), \\
    CC(s,i)&=&pf\eta_s\eta_i+ACC(s,i), \\
    ACC(s,i)&=&f\frac{p\eta_sSC(s)}{SC(s)_{net}}\frac{p\eta_iSC(i)}{SC(i)_{net}}+O(p^3).
	\end{eqnarray}
    Note that the detection efficiency of two photons at one detector is $1-\left(1-\eta_{s/i}\right)^2\approx 2\eta_{s/i}$; accidentals coincidence $ACC(s,i)$ is complex since one photon from dual-pump SFWM process and another from single-pump SFWM process also give a twofold coincidence. Assuming dark counts $DC_{s/i}$ is similar, we fit the experimental data [Tab.\ref{tab:SFWM}] and estimate the pair-generation rate $p$ of $\sim$0.003 per pulse, the losses $\eta_s$ of 15 dB, and $\eta_i$ of 16 dB.

    Next, we consider the four-photon component,$1/2\,g^2(z_0^+ z_1^+ )\otimes(z_0^+ z_1^+)\left|\rm{vac}\right>=g^2\left|2_0,2_1\right>$.
    The generation rate of two pairs is $g^4=p^2$. From these generated four photons, the postselection efficiency to get fourfold coincidence is 3/32, which has contained the one-by-four splitting loss in the Dicke-state projection. So, the four-photon coincidence rate without multi-pair noise can be estimated by,
    \begin{eqnarray}
        CC_{4-photon\ state}&=&4^4 \frac{3}{32}p^2 f\eta_A\eta_B\eta_C\eta_D \nonumber\\
        &=&4^4 \frac{3}{32}\frac{\left(CC_{A0,B1}-ACC_{A0,B1}\right)\left(CC_{C0,D1}-ACC_{C0,D1}\right)}{f}.
    \end{eqnarray}
    We can estimate the four-photon coincidence rate from pair-generation rate and detection loss or directly from the measured two-photon coincidence. The two-photon coincidence measured in tomography is a bit higher of 1100-1600 Hz floating with power fluctuation of pump. Thus, the estimated total four-photon coincidence rate is 200-440 per hour. The measured total four-photon coincidence counts are $\sim$960 per setting in 2.8 hours, which matches well with the estimation.

    \section{Spectral purity of DMZI-ring photon-pair sources}
\begin{figure}[htbp]
		\includegraphics[width=0.3\textwidth]{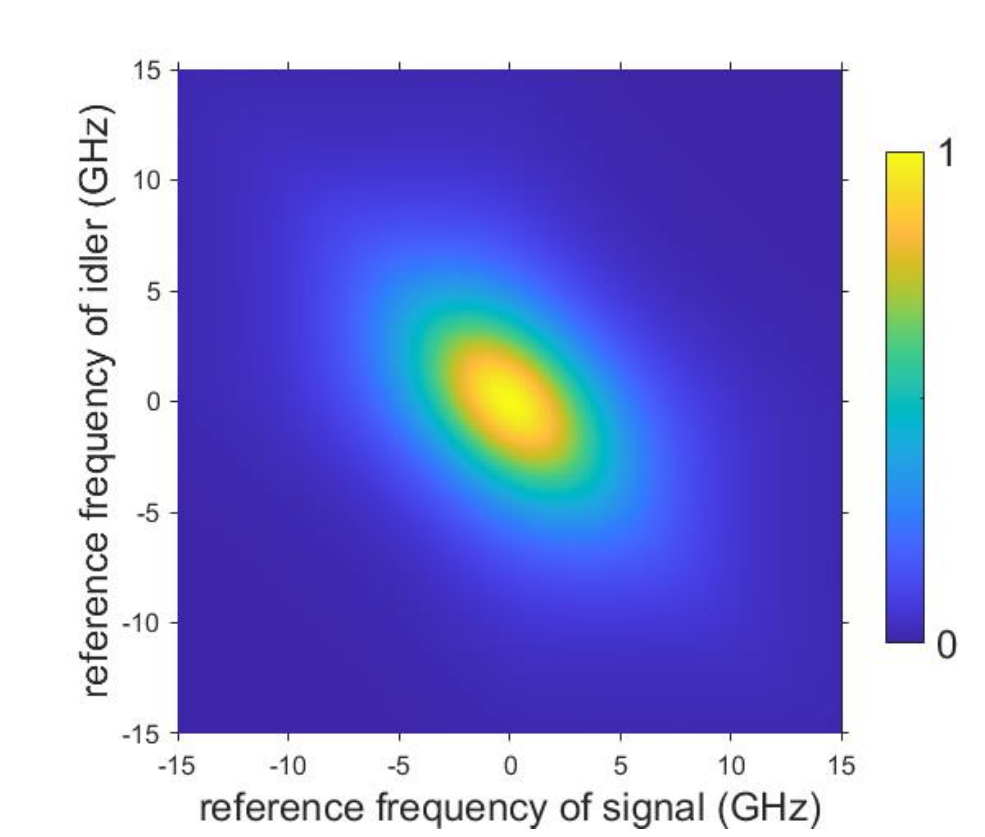}
		\caption{\label{fig:siadd3} Simulated joint spectral intensity (JSI) of the DMZI-ring sources. It shows a purity of 83\% for our design.}
	\end{figure}

    We estimate the spectral purity of photon-pair sources via the joint spectral intensity (JSI) from simulation. The spectral purity, $Pur.$, or the Schmidt number $K=1/Pur.$ quantifies the frequency correlation between signal and idler photons. The spectral purity of the silicon DMZI-ring sources has been explicitly analyzed in the Ref.\cite{vernon2017truly,liu2020high}, showing that a higher spectral purity can be achieved by increasing the ratio $Q_{s/i}/Q_p$ between Q factor of signal (idler) and pump. One can achieve a purity of $\sim$93\% for a single-bus ring resonator ($Q_{s/i}/Q_p=1$). For our DMZI-rings, we purposefully design overcoupling conditions for signal/idler photons to get a low photon-extraction loss and a high measured coincidence rate. The ratio $Q_{s/i}/Q_p$ is $\sim$0.5.  We simulated the JSI of our DMZI-ring sources [Fig.\ref{fig:siadd3}] and obtained a purity of $\sim$0.83.
    
	\section{Analysis on the visibility of two-photon interference}
\begin{figure}[htbp]
		\includegraphics[width=0.9\textwidth]{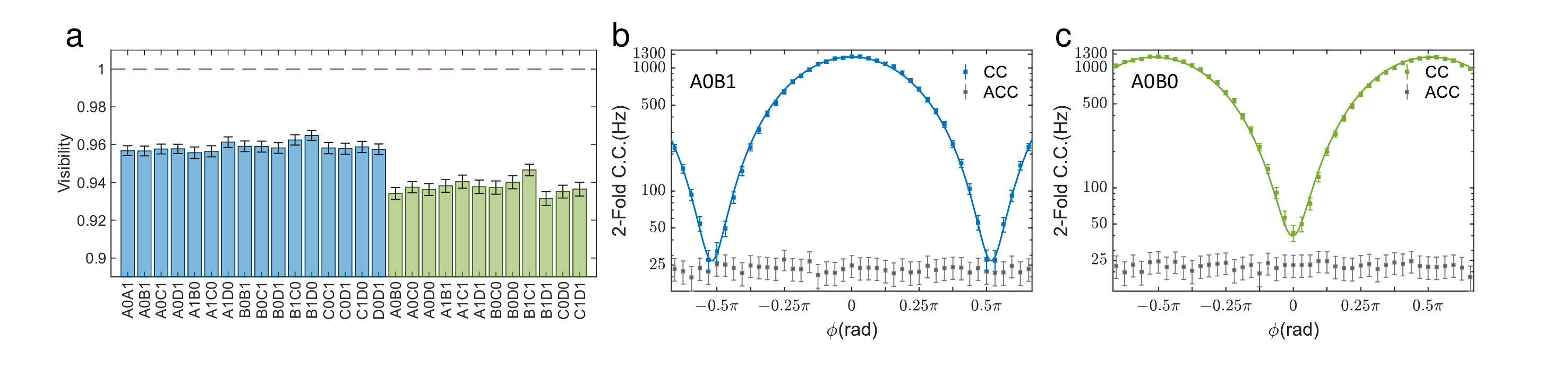}
		\caption{(a) Visibilities of all 28 two-photon interference fringes. Coincidence between photons in different path modes corresponds to $\left|\Psi^+\right>$ state (blue bars); Coincidence between photons in the same path mode corresponds to $\left|\Phi^-\right>$ state (green bars). (b) Interference fringe of $\left|\Psi^+\right>$ detected between A0 and B1. (c) Interference fringe of $\left|\Phi^-\right>$ detected between A0 and B0. The measured accidental coincidence counts between different pulse times are shown in gray.}
        \label{fig:si5} 
	\end{figure}

	In this section, we analyze the visibility of two-photon interference. In the main text, the measurement on $\left|\Psi^+\right>$ and $\left|\Phi^-\right>$ shows complementary fringes. Here we display visibilities of 28 fringes from all possible twofold coincidences of eight outputs [Fig.\ref{fig:si5}(a)], in which 16 correspond to $\left|\Psi^+\right>$ state and 12 correspond to $\left|\Phi^-\right>$ state. The average raw visibilities are $95.87\%\pm0.07\%$ for $\left|\Psi^+\right>$ state and $93.76\%\pm0.09\%$ for $\left|\Phi^-\right>$ state. 
 
    We conjecture the lower visibility of $\left|\Phi^-\right>$ state is mainly caused by higher accidental coincidence. The accidental coincidence is no longer the same as the measured accidental coincidence between different pulse times. The ratio between these two accidental coincidence counts is related to the sources' spectral purity $Pur.$ and the mixture of three kinds of SFWM processes.

\begin{figure}[htbp]
		\includegraphics[width=0.6\textwidth]{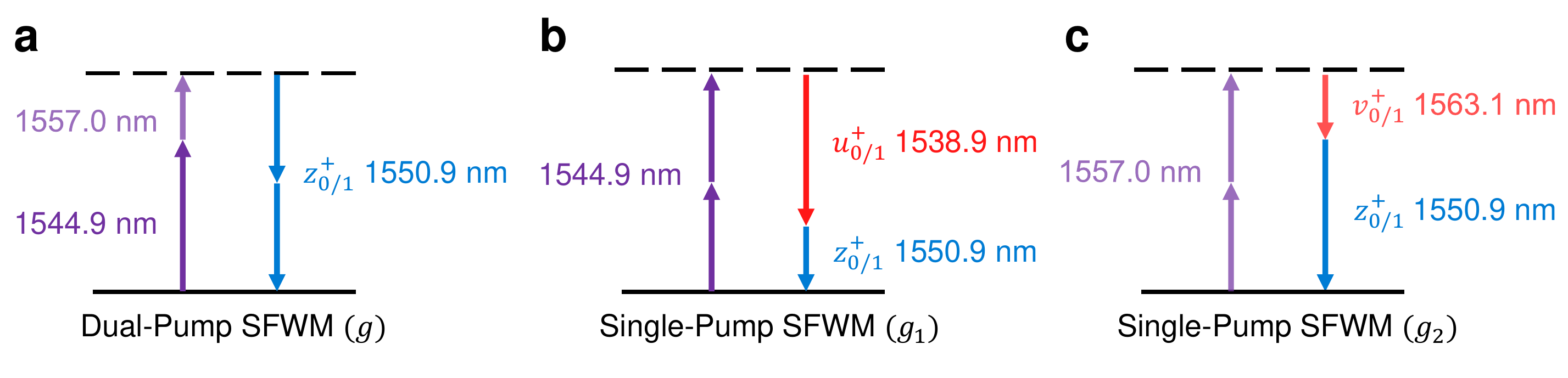}
		\caption{Three SFWM processes under dual-color pulsed pump. (a) Dual-pump SFWM process produces frequency-degenerate photons (1550.9 nm). (b) Single pump at 1544.9 nm produces two nondegenerate photons (1550.9 nm and 1538.9 nm). (c) Single pump at 1557.0 nm produces photons (1550.9 nm and 1563.1 nm).}
        \label{fig:siadd4} 
	\end{figure}

    Considering the impurity of source and all kinds of SFWM processes, the whole photon state after RHOM interference can be approximately written as,
    \begin{eqnarray}
        \left|\Psi_2(\phi=0)\right> &=& 
        \exp{\left[-gZ_0^+Z_1^+ -\frac{1}{\sqrt{2}}g_1\left(Z_0^+U_1^+ + Z_1^+U_0^+\right)-\frac{1}{\sqrt{2}}g_2\left(Z_0^+V_1^+ + Z_1^+V_0^+\right) \right]}\left|\rm{vac}\right>, \label{eq:rhom-1}\\
        \left|\Psi_2(\phi=\frac{\pi}{2})\right> &=& 
        \exp{\left[-\frac{1}{2}g\left(Z_0^+Z_0^+ - Z_1^+Z_1^+\right)-\frac{1}{\sqrt{2}}g_1\left(Z_0^+U_0^+ - Z_1^+U_1^+\right)-\frac{1}{\sqrt{2}}g_2\left(Z_0^+V_0^+ - Z_1^+V_1^+\right) \right]}\left|\rm{vac}\right>.\label{eq:rhom-2}
    \end{eqnarray}
    Note that $Z_0^+ X_1^+=\Sigma \lambda_i \left(z_0^+\right)_i \left(x_1^+\right)_i $ with $\Sigma \lambda_i^2=1, X=Z,U,V$. $U_{0/1}^+$ and $V_{0/1}^+$ are the creation operators of photons with another frequency from single-pump SFWM processes [Fig. \ref{fig:siadd4}]. $g$, $g_1$ and $g_2$ correspond to non-linear coupling of the three SFWM processes. Here, we assume the same distribution of joint spectral function in different SFWM processes due to the similar pump and signal/idler bandwidths. The purity $Pur.$ can be quantified with the Schmidt number $K$ defined in \cite{christ2011probing,grobe1994measure} as,
    \begin{equation}
        Pur.=1/K=\Sigma \lambda_i^4.
    \end{equation}
    
	After RHOM interference, photons are split into two ports A and B [Eq.\ref{eq:S1-5} and \ref{eq:S1-6}], collected, and detected with efficiencies $\eta_A$ and $\eta_B$. The measured accidental coincidence between different pulses can be estimated with the coincidence of two single-count rates per pulse,
    \begin{eqnarray}
        SC_A&\sim&\frac{1}{2}g^2\eta_A+\frac{1}{4}g_1^2\eta_A+\frac{1}{4}g_2^2\eta_A,\\
        SC_B&\sim&\frac{1}{2}g^2\eta_B+\frac{1}{4}g_1^2\eta_B+\frac{1}{4}g_2^2\eta_B,\\
        ACC_{measured}&=&SC_A\cdot SC_B=\frac{1}{4}\left(g^2+\frac{1}{2}g_1^2+\frac{1}{2}g_2^2\right)^2\eta_A\eta_B.
    \end{eqnarray}
    Given that the single-count rate remains constant, we get a similar count rate of measured accidental coincidence when scanning the phase $\phi$ [Fig.\ref{fig:si5}(b) and (c)].

    The accidental coincidence in the minimum point of the fringes influences the measured visibilities. It can be estimated via the two-pair components. For the minimum point of the fringes of $\left|\Psi^+\right>$ with $\phi=\frac{\pi}{2}$, we calculate the accidental coincidence between A0 and B1,
    \begin{equation}
        ACC_{\Psi^+ min.}=\frac{1}{4}\eta_A\eta_B\left(g^4+\frac{1}{4}g_1^4+\frac{1}{4}g_2^4+g^2 g_1^2+g^2 g_2^2+\frac{1}{2}g_1^2 g_2^2 \right)=ACC_{measured}.  
    \end{equation}    
    We can see the accidental coincidence is the same as the measured one. The measured minimum coincidence of $\left|\Psi^+\right>$ is similar to the measured accidental coincidence [Fig.\ref{fig:si5}(b)], resulting in high net visibility close to 100\%. It shows high indistinguishability between the emitted photons from two DMZI-rings.

    However, for the minimum point of the fringes of $\left|\Phi^-\right>$ with $\phi=0$, frequency-degenerate photons make difference. The accidental coincidence contains the contribution of two pairs from the same or different SFWM processes:
    \begin{eqnarray}
        \rm{Same\ SFWM:} &\frac{1}{2}\left(Z_0^+X_1^+\right)^2=\frac{1}{2}\sum\limits_i\lambda_i^2(z_0^+)_i^2(x_1^+)_i^2+\sum\limits_{i<j}\lambda_i\lambda_j(z_0^+)_i(z_0^+)_j(x_1^+)_i(x_1^+)_j,\\
        \rm{Different\ SFWM:} & Z_0^+X_1^+Z_0^+Y_1^+=\sum\limits_i\lambda_i^2(z_0^+)_i^2(x_1^+)_i(y_1^+)_i+\sum\limits_{i\neq j}\lambda_i\lambda_j(z_0^+)_i(z_0^+)_j(x_1^+)_i(y_1^+)_j.
    \end{eqnarray}
    Either $(z_0^+)_i^2$ or $(z_0^+)_i(z_0^+)_j$ gives two photons for twofold coincidence detection. The possibility they split into two ports is 1/2 because of the exchange symmetry, while it is 1/4 for two photons in different path modes. Thus, their contributions to the accidental coincidence between A0 and B0 can be estimated,
    \begin{eqnarray}
        ACC_{Same.}=g^4 \left(\sum\limits_i\lambda_i^4+ \sum\limits_{i<j}\lambda_i^2\lambda_j^2\right)\frac{1}{2}\eta_A\eta_B=\frac{1}{4}g^4\eta_A\eta_B\left(\sum\limits_i\lambda_i^4+ \sum\limits_{i}\lambda_i^2\sum\limits_{j}\lambda_j^2\right)=\frac{1+Pur.}{4}g^4\eta_A\eta_B,\\
        ACC_{Dif.}=g^2g_1^2\left(2\sum\limits_i\lambda_i^4+ \sum\limits_{i\neq j}\lambda_i^2\lambda_j^2\right)\frac{1}{2}\eta_A\eta_B =\frac{1}{2}g^4\eta_A\eta_B\left(\sum\limits_i\lambda_i^4+ \sum\limits_{i}\lambda_i^2\sum\limits_{j}\lambda_j^2\right) =\frac{1+Pur.}{2}g^2g_1^2\eta_A\eta_B.
    \end{eqnarray}
    Considering all possibilities from the same or different SFWM processes, we get the accidental coincidence at the minimum point of the fringes of $\left|\Phi^-\right>$:
    \begin{equation}
        ACC_{\Phi^- min.}=\frac{1+Pur.}{4}\eta_A\eta_B\left(g^2+\frac{1}{2}g_1^2+\frac{1}{2}g_2^2\right)^2=(1+Pur.)ACC_{measured}.   
    \end{equation}  
    The accidental coincidence of $\left|\Phi^-\right>$ is higher than the measured one, resulting in the larger minimum [Fig.\ref{fig:si5}(c)] and the lower visibility. The ratio between these two accidental coincidence counts quantifies the unheralded second-order correlation $g^{(2)}$,
    \begin{equation}
        ratio=g^{(2)}=1+Pur.
    \end{equation}

    In the experiment, the average ratio of the minimum coincidence and measured accidental coincidence of $\left|\Phi^-\right>$ is $1.786\pm0.049$, which is consistent with the simulated purity of the DMZI-rings. The large deviation is mainly caused by the low measured counts. This measurement is from the interference of two DMZI-rings, which differs from $g^{(2)}$ test of a single source. The difference in the joint spectral function of three SFWM processes may also have some influence on the measured ratio.
 
 
    \section{The purity of the two-photon and four-photon entangled states}
    The purity of an entangled state can be obtained by $P=\rm{Tr}\left(\rho_{\text{exp}}\rho_{\text{exp}}\right)$, which quantifies how much a state is mixed. We perform quantum tomography for each two-photon state between two of ports A-D under the same pump condition as generating the four-photon state. By applying $\phi=0$, the output two-photon state is $\left|\Psi^+\right>$. The fidelities and the purities of two-photon states are listed in Tab.\ref{tab:pur2photon}. The average fidelity is $0.942\pm0.001$, which gives an average purity of $0.907\pm0.001$.

    For the generated four-photon state, we obtain the purities of $0.710\pm0.004$ for $\left|\Psi_4(0) \right>$ and $0.710\pm0.004$ for $\left|\Psi_4(\pi/2) \right>$, respectively. The moderate fidelity and purity are due to the non-negligible accidental coincidence and higher-order emissions with the high pump power.
    
    \begin{table}[htbp]
        \centering
        \begin{tabular}{ccc}
        \hline
         Photons &  Fidelity to $\left|\Psi^+\right>$ & Purity\\
        \hline
        (A,B) & $0.945\pm0.001$ &   $0.906\pm0.001$\\
        (A,C) & $0.942\pm0.001$ &   $0.905\pm0.001$\\
        (A,D) & $0.934\pm0.001$ &   $0.915\pm0.001$\\
        (B,C) & $0.946\pm0.001$ &   $0.902\pm0.001$\\
        (B,D) & $0.945\pm0.001$ &   $0.908\pm0.001$\\
        (C,D) & $0.940\pm0.001$ &   $0.908\pm0.001$\\
        \hline
        \end{tabular}
        \caption{Fidelities and Purities of two-photon entangled states.}
        \label{tab:pur2photon}
    \end{table}

	\section{The imaginary part of density matrixes of two states from tomography}
\begin{figure}[htbp]
		\includegraphics[width=0.6\textwidth]{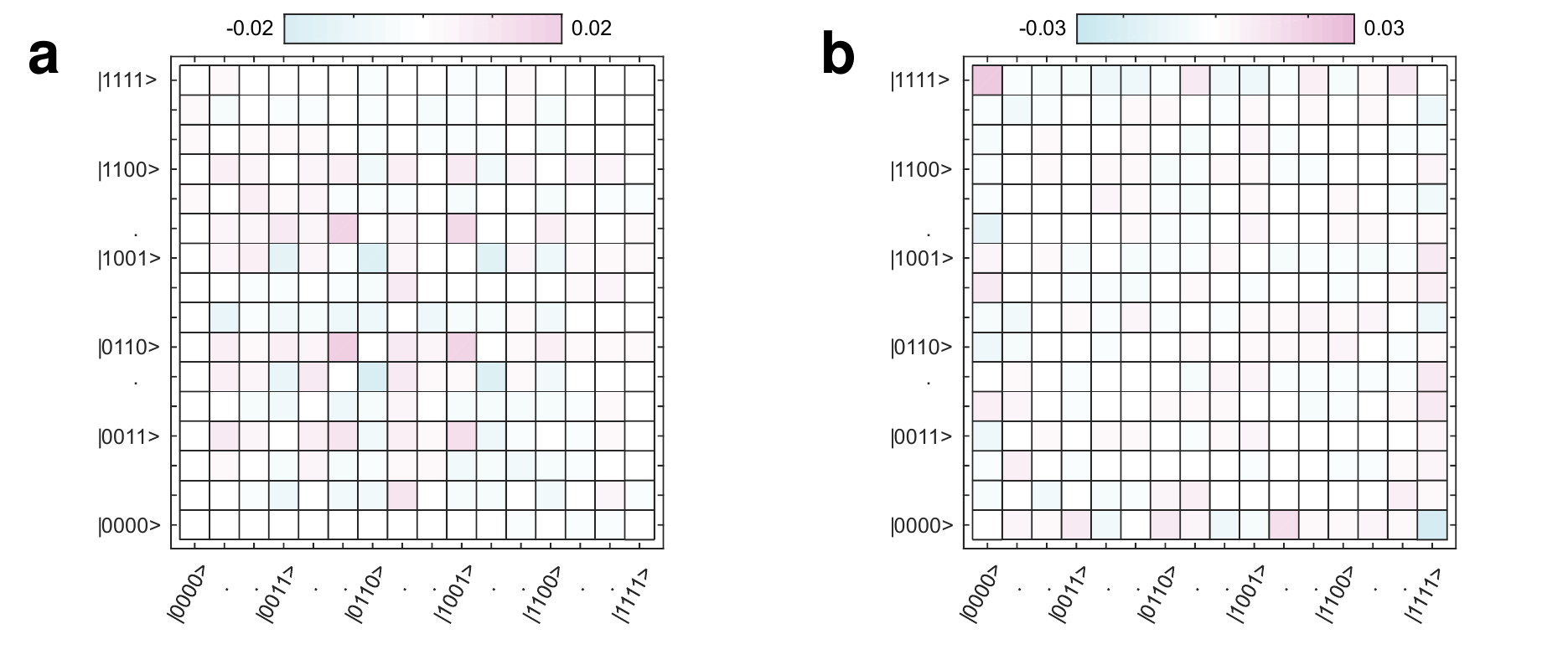}
		\caption{\label{fig:si6} The difference between the imagine part of the density matrix and the ideal case for different states. (a) The symmetric Dicke state $\left|D_4^2\right>$ by setting $\phi=0.$ (b) The superposition of four-photon GHZ state $\left|GHZ_4\right>$ and symmetric Dicke state $\left|D_4^2\right>$ by setting $\phi=\pi/2$.}
	\end{figure}

	We show both the real [Fig.4 in the main text] and imaginary parts [Fig.\ref{fig:si6}] of density matrixes from tomography on two states $\left|\Psi\left(0\right)\right>$ and $\left|\Psi\left(\pi/2\right)\right>$. The imaginary part of the ideal density matrix is zero. We see the measured ones are fairly small with the difference to the ideal case smaller than 0.03.
	
	\section{Conversion between different classes of three-photon entangled states}
\begin{figure}[htbp]
		\includegraphics[width=0.6\textwidth]{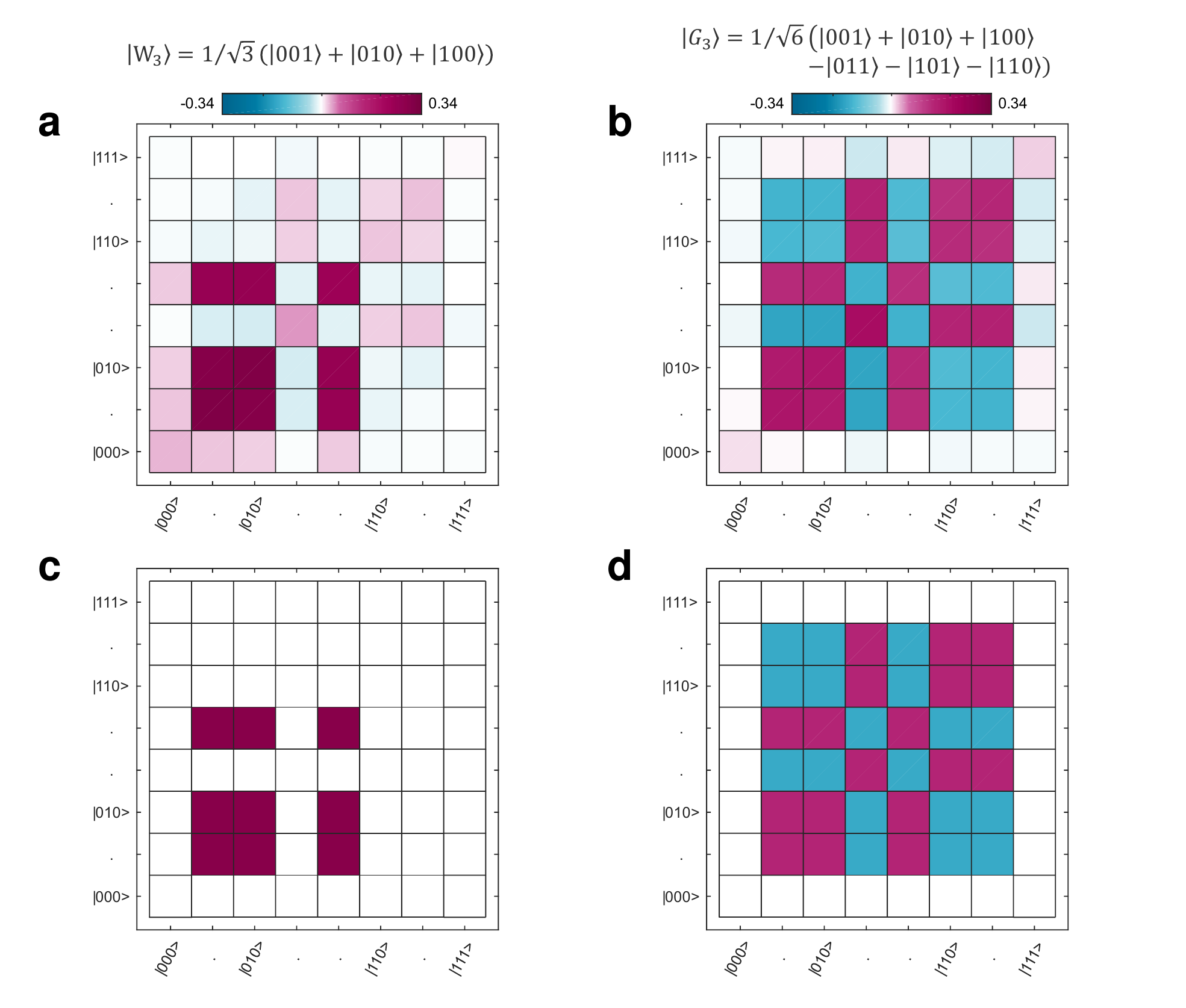}
		\caption{\label{fig:si7} The density matrix for three-photon entanged states. (a) The three-photon W state  $\left|W_3\right>$ by projecting a photon of  $\left|D_4^2\right>$ into mode  $\left|1\right>$. (b) The three-photon G3 state  $\left|G_3\right>$. by projecting a photon of $\left|D_4^2\right>$ into mode $\left|-\right>$. The experimentally obtained quantum-state fidelities are $0.840\pm0.003$ and $0.854\pm0.003$, respectively. The ideal density matrix are shown in (c) and (d), for comparison.}
	\end{figure}

	Employing a projecting measurement on a multiphoton Dicke state, we can obtain different entangled states with less photons \cite{kiesel2007experimental}. As for the symmetric Dicke state, projecting one of photons in mode $\left|1\right>$, we get a so-called W-state: $\left|W_3\right>=1/\sqrt3\ \left(\left|001\right>+\left|010\right>+\left|100\right>\right)$, 
	which is the three-photon Dicke state with one excitation. By projecting one photon in mode $\left|-\right>$, we get a so-called G3 state \cite{sen2003multiqubit}: $\left|G_3\right>=1/\sqrt6\ \left(\left|001\right>+\left|010\right>+\left|100\right>-\left|011\right>-\left|101\right>-\left|110\right>\right)$, which belongs to the GHZ class \cite{kiesel2007experimental}. Experimentally, we observe the  $\left|W_3\right>$ state with a fidelity of $0.840\pm0.003$ and the $\left|G_3\right>$ state with a fidelity of $0.854\pm0.003$. The measured density matrixes are shown in Fig.\ref{fig:si7}. These states with high fidelities show various multiphoton entangled states can be generated on the same chip with proper control and measurement, indicating the high quality and universality of our multiphoton source.

	\section{Maximal singlet fraction of four-photon states}
\begin{figure}[htbp]
		\includegraphics[width=0.8\textwidth]{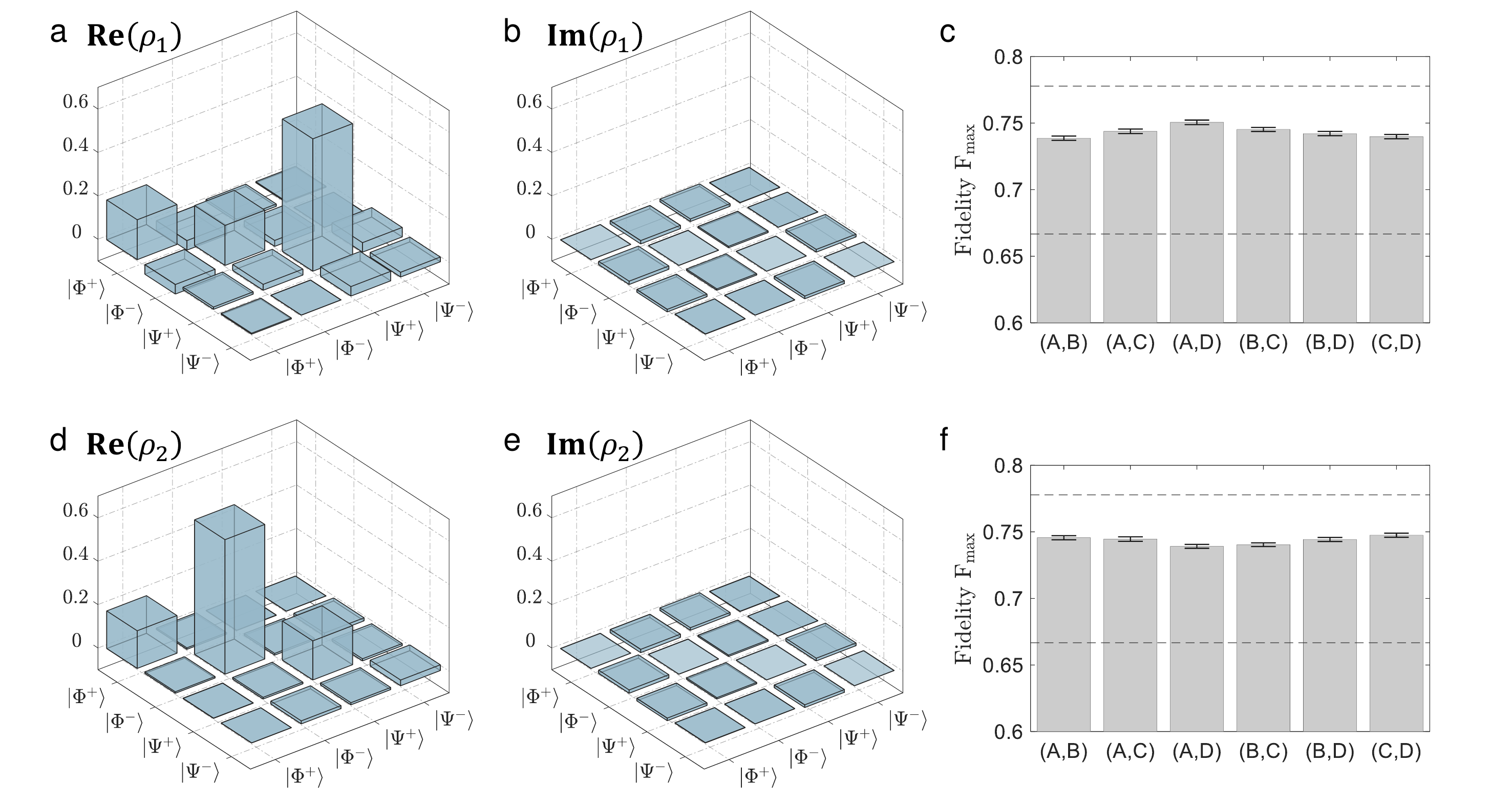}
		\caption{\label{fig:si8} Singlet fraction of two-photon state. (a) and (b) Real and imagine part of the density matrix for the two-photon state $\rho_1$ between photon A and B traced from $\left|\Psi\left(0\right)\right>$. The two-photon state between other pair of photons is similar due to the symmetry of the four-photon state. (c) Maximal achievable fidelities of teleportation $F_{max}$ for all pairs of photons. The upper and lower dash line show the ideal value and the classical limit. (d), (e), and (f) show the case for $\left|\Psi\left(\pi/2\right)\right>$.}
	\end{figure}

	The maximal singlet fraction $F_{msf}$ is one of the entanglement measures for two entangled qudits. Tracing two photons from the Dicke state, we calculate $F_{msf}$ of any photon pair from the four-photon state. Theoretically, tracing two photons from state $\left|\Psi\left(0\right)\right>$, we get the two-photon state $\rho_1=1/6\ \left|\Phi^+\right>\left<\Phi^+\right|+1/6\ \left|\Phi^-\right>\left<\Phi^-\right|+2/3\ \left|\Psi^+\right>\left<\Psi^+\right|$; tracing two photons from state $\left|\Psi\left(\pi/2\right)\right>$, we get the two-photon state $\rho_2=1/6\ \left|\Phi^+\right>\left<\Phi^+\right|+2/3\ \left|\Phi^-\right>\left<\Phi^-\right|+1/6\ \left|\Psi^+\right>\left<\Psi^+\right|$. 
	Figure \ref{fig:si8} (a, b, d, and e) show the real and imagine part of density matrixes of $\rho_1$ and $\rho_2$, respectively, which are rotated into Bell basis and agree well with the theory. The maximal singlet fraction $F_{msf}$ is given by the maximal overlap of the state with a maximally entangled state under local operations. Since the two-photon mixed states are diagonalizable in Bell basis, $F_{msf}$ is equal to the maximal proportion 2/3 of one Bell-state component. Experimentally, we measured a mean maximal singlet fraction $\sim0.614\pm0.001$ for $\left|\Psi\left(0\right)\right>$ and  $\sim0.616\pm0.001$ for $\left|\Psi\left(\pi/2\right)\right>$, indicating high entanglement persistency against loss of photons.
	
	As any photon pair still remain high entanglement, the generated four-photon state has the potential in quantum networking tasks, such as teleportation and telecloning \cite{murao1999quantum}.
	We can further estimate the maximal achievable fidelity of teleportation $F_{max}$ via the relation $F_{max}=\left(F_{msf}d+1\right)/\left(d+1\right)$ \cite{horodecki1999general}.
	Note that $d$ is the dimension of photons and $d=2$ for qubits. It is worthwhile to add that the optimal cloning fidelity for symmetric $1\rightarrow M$ telecloning tasks is the exactly same as $F_{max}$ \cite{gisin1997optimal,prevedel2009experimental}. Here, we display $F_{max}$ for all pairs of photons by tracing the rest of photons of states $\left|\Psi\left(0\right)\right>$ and $\left|\Psi\left(\pi/2\right)\right>$ [Fig.\ref{fig:si8} (c) and (f) respectively]. The ideal value is 0.78 and the classical limit is 0.67. In our cases of four-photon state, we get a mean maximal fidelity $\sim0.743\pm0.001$ for $\left|\Psi\left(0\right)\right>$ and  $\sim0.744\pm0.001$ for $\left|\Psi\left(\pi/2\right)\right>$.
	
	\section{Four-photon states in the rotated basis}
\begin{figure}[htbp]
		\includegraphics[width=0.6\textwidth]{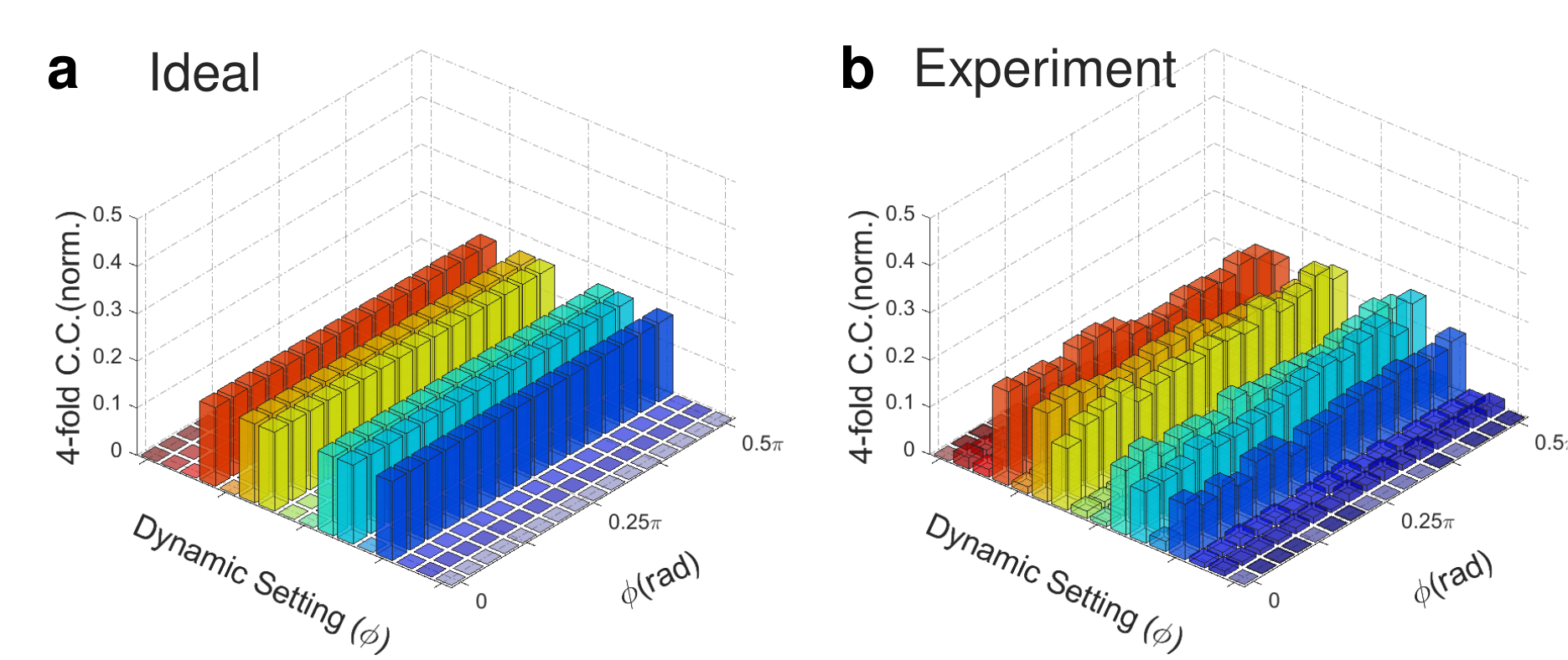}
		\caption{\label{fig:statecontroltheta} Collectively coherent control of the four-photon state. (a) and (b), the expected ideal and experimentally measured coincidence distributions for photons measured in the rotated basis of $\left\{\left|\theta\right>,\left|\theta_\perp\right>\right\}$. The distribution remains constant.}
	\end{figure}

	We rewrite the four-photon state into the rotated basis $\left\{\left|\theta\right>,\left|\theta_\perp\right>\right\}=\left\{\left(\begin{matrix}\sin{\phi/2}\\\cos{\phi/2}\end{matrix}\right),\left(\begin{matrix}\cos{\phi/2}\\-\sin{\phi/2}\end{matrix}\right) \right\}$, which is dynamic changing according to the phase $\phi$, and obtain:
	\begin{equation}
		\left|\Psi_4\left(\phi\right)\right>=\frac{1}{\sqrt6}
		\left(\left|\theta \theta \theta_\perp \theta_\perp \right>
		+\left|\theta \theta_\perp \theta \theta_\perp \right>+\left|\theta \theta_\perp \theta_\perp \theta \right>+\left|\theta_\perp \theta \theta \theta_\perp \right>+\left|\theta_\perp \theta \theta_\perp \theta \right>
		+\left|\theta_\perp \theta_\perp \theta \theta \right>\right).
		\label{eq:4photonstatetheta}
	\end{equation}
    We can see the state is rotated back to the symmetric four-photon Dicke state. The theoretical and the experimental results are shown in Fig.\ref{fig:statecontroltheta}, respectively, which agree well with each other. It indicates that the phase $\phi$ can collectively rotate all four photons with the same degrees at the same time through quantum interference. Tuning the relative phase $\phi$ is equivalent to imposing the same phase shift on each MZI and rotates the four-photon state as a whole.
	
\bibliography{Dicke_bib}

\end{document}